\documentclass[10pt]{article}
\usepackage{a4}
\usepackage{graphicx}
\usepackage{color}
\usepackage{cite}
\usepackage{epsfig}
\usepackage{amsfonts}
\begin{document}
\begin{center}
\title[ \textbf{ON REAL INTRINSIC WALL CROSSINGS}\\
\vspace{0.5cm} {{\bf  Stefano Bellucci\footnote{e-mail:
bellucci@lnf.infn.it }$^{,a}$ and Bhupendra Nath
Tiwari\footnote{e-mail: tiwari@lnf.infn.it}$^{,a}$}}
\end{center}
\vspace{0.3cm}
\begin{center}
{$^a$ \it INFN- Laboratori Nazionali di Frascati,\\
 Via E. Fermi 40, 00044, Frascati, Italy}\\
\end{center}
\vspace{0.3cm} We study moduli space stabilization of a class of
BPS configurations from the perspective of the real intrinsic
Riemannian geometry. Our analysis exhibits a set of implications
towards the stability of the D-term potentials, defined for a set
of abelian scalar fields. In particular, we show that the nature
of marginal and threshold walls of stabilities may be investigated
by real geometric methods. Interestingly, we find that the leading
order contributions may easily be accomplished by translations of
the Fayet parameter. Specifically, we notice that the various
possible linear, planar, hyper-planar and the entire moduli space
stabilities may easily be reduced to certain polynomials in the
Fayet parameter. For a set of finitely many real scalar fields, it
may be further inferred that the intrinsic scalar curvature
defines the global nature and range of vacuum correlations.
Whereas, the underlying moduli space configuration corresponds to
a non-interacting basis at the zeros of the scalar curvature,
where the scalar fields become un-correlated. The divergences of
the scalar curvature provide possible phase structures, viz., wall
of stability, phase transition, if any, in the chosen moduli
configuration. The present analysis opens up a
new avenue towards the stabilizations of gauge and string moduli.

\newpage

\begin{Large} \textbf{Contents:} \end{Large}
\begin{enumerate}
\item{Introduction.}
\item{D-term Potentials and Real Intrinsic Geometry.}
\item{Marginal Stability:}
\subitem{$3.1$\ \ Single Complex Scalar.} \subitem{$3.2$ \ \
Multiple Complex Scalars.}
\item{Threshold Stability:}
\subitem{$4.1$\ \ Simple Complex Scalars.} \subitem{$4.2$ \ \
General Complex scalars.}
\item{Conclusion and Remarks.}
\end{enumerate}
\section{Introduction}
The moduli Space geometry is a cornerstone of modern string theory
compactifications and divulges the nature of a class of intriguing
quantum field theory, supersymmetric gauge theory configurations
and string theory dynamics in all possible dimensions
\cite{1a,1b}. In fact, it has been a long standing problem how to
resolve the fundamental issue of moduli stabilization in
cosmology, gauge theory and superstring theory
\cite{2a,2b,2c,2d,2e,2f,2g,2h,2i,2l,2m}. Moreover, the problem
gets a more interesting facet from the perspective of moduli
stabilization and flux compactifications \cite{3a,3b,3c}, which
are a class of configurations involving the D-terms contributions.
Nevertheless, our analysis leads to an interesting conclusion for
the notion of the moduli stability of underlying configurations.
Such an introduction of the Fayet model \cite{4a,4b} captures the
basic leading order features of the moduli stabilization problem.
The promising application of the present analysis concerns certain
decaying supergravity configurations. Herewith, we shall consider
the intrinsic real geometric framework to analyze the phenomena of
wall crossings, and in particular, we shall focus our attention on
the $D$-term potentials.

It is worth mentioning that the charged extremal black holes in $D
= 4$, $ N = 2$ supergravity may be characterized by a set of
electric-magnetic charges $\{q_J, p^I\}$, which arise from the
flux integrals of the field strength tensor and the corresponding
Poincar\'{e} dual. On the other hand, the scalar fields arising
from the compactification of both the string theory and M-theory
yield a set of moduli fields, which in effect parameterize a
certain compact internal space. The extremal charged black hole
solutions can be viewed as BPS solitons, which interpolate between
the asymptotic infinity and $AdS_2 \times S^2$ near horizon
geometry \cite{5,6}. The spherical symmetry, in turn, determines
the underlying interpolation as the radial evolution of the scalar
moduli, which encodes the consequent changes of the underlying
compact internal manifold. Moreover, one finds that there exists
the flat Minkowskian manifold at the asymptotic infinity, for a
given scalar moduli configuration. The asymptotic ADM mass, as an
arbitrary parameter, is described by the complex central extension
$Z_{\infty}$ of the underlying supersymmetry algebra \cite{5,6}.
Subsequently, it turns out that the ADM mass satisfies $M (p, q,
\varphi^a) = | Z_{\infty} |$.

In such cases, the near horizon geometry of an arbitrary extremal
black hole reduces to the $AdS_2 \times S^2$ manifold, which
describes the underlying Bertotti-Robinson vacuum. The area of the
black hole horizon $A$, and hence its macroscopic entropy is given
as $S_{macro}= \pi |Z_{\infty}|^2$. However, it turns out that the
radial variation of the moduli is described by a damped geodesic
equation, which flows to an attractive fixed point at the horizon.
Thus, it may solely be determined by the charges carried out by
the chosen extremal black hole. Such attractors \cite{7} have been
further studied from the perspective of the criticality of the
black hole effective potential in $D = 4$, $ N = 2$ supergravities
coupled to $n_V$ abelian vector multiplets. An asymptotically flat
extremal black hole background may in turn be described by $(2n_V
+ 2)$-dyonic charges and $n_V$-complex scalar fields. Thus, the
scalar moduli parameterize a $n_V$-dimensional special K\"{a}hler
manifold, see \cite{7} and references therein.

From the perspective of the attractor horizon geometries, the
study of supergravity theories led to the fact that an extremal
dyonic black hole has a non-vanishing Bekenstein-Hawking entropy,
see for example \cite{8}. In Ref. \cite{8}, such an analysis has
been given for 1/2-BPS and non-BPS (non-supersymmetric) attractors
with a non-vanishing central charge, in the context of $D = 4$, $
\mathcal N= 2$ ungauged supergravity coupled to a $n_V$-number of
abelian vector multiplets. Further, this leads to an interesting
classification of the (i) orbits of the classical U-duality group,
in the symplectic representation, and (ii) moduli spaces
associated with the non-BPS attractors, in the context of
symmetric special K\"{a}hler geometries. From the perspective of
the present analysis, the case of non-extremal black brane
configurations may be similarly considered by adding the
corresponding anti-branes to the extremal black brane
configurations. Thus, the computation of the associated black
brane entropy may be performed in both the microscopic and the
macroscopic descriptions, viz., \cite{9,10}. In fact,
Refs.\cite{9,10} show that such a consideration leads to a
perturbative matching between the $S_{micro}$ and $S_{macro}$, for
a set of brane charges and a given mass. Interestingly, the
addition of the mass to an extremal configuration defines a
nonextremal configuration.

From the perspective of thermodynamic geometry, it turns out that
the problem involved could be examined at the attractor fixed
point(s), where the moduli fields are stabilized via the attractor
equations \cite{11a,11b,11c,11d,11e,11f,11g,11h,11i,11j,11k}, and
thus expressed in terms of the invariant charges of the
configuration. The promising geometry, thus arising at an
attractor fixed point, describes thermodynamic fluctuations in the
entropy of the underlying black hole and a class of associated
configurations
\cite{12a,12b,12c,12d,12e,12f,12g,12h,12i,12j,12k,12l,12m,12n,
12o,12oo,12ooo,12o1,12o2,12o3,12p,12p1,12p2,12p3,12q,12r,12s,12t,12u}.
Furthermore, some interesting directions have been explored for
the non-extreme Calabi-Yau configurations
\cite{13a,13b,13c,13d,13e,13f}. In this direction, one may
investigate the moduli stabilization problem further, from the
perspective of \cite{14, 15, 16}. In particular, one finds, for a
given K\"{a}hler potential as a function of the moduli, that the
case of the extreme Calabi-Yau manifolds corresponds to certain
simplifications in the computation of the attractor flow
equations. Interestingly, the intrinsic geometric transformations
turn out to be extraordinarily informative, and in particular,
such transformations uprise to certain symmetries of the effective
potential of the theory. In such considerations, it is worth
mentioning that the extremization problem involves an appropriate
notion of the covariant derivative, and thus one needs to
incorporate the contributions coming from the central charge of
the theory \cite{17a,17b,17c}, which offers appropriate setups for
the general K\"{a}hler moduli configurations.

Nevertheless, the framework of the present paper outlines an
appreciation of the extremization problem. Consequently, one may
consider the examinations from the stability of the moduli
configurations. For the string theory motivated backgrounds, such
a system could be obtained by an appropriate compactification of a
higher dimensional configuration. For a given finite dimensional
moduli system, one may consider the Hessian of the effective
potential and thereby employ the notion of the intrinsic geometry.
In particular, it may be anticipated that such an investigation
would correspond to a unified depiction of both the statistical
fluctuations and intrinsic Riemannian geometric stabilities. Thus,
it may be expected that there exists a certain geometric notion of
the stability, for a given (black hole) effective potential. The
classical configurations effectively reduce to an underlying
thermodynamic system, at the attractor fixed point(s). From the
viewpoint of the present consideration, there exists a class of
black hole solutions \cite{18a,18b,18c,18d,18e,18f}, leading to an
interesting attractor flow and the corresponding thermodynamic
configurations. Furthermore, there have been considerations of
incorporating higher order corrections \cite{19a,19b}, from the
perspective of the string theory. In general, the notion of
stability may be divulged from the perspective of the moduli flow.
Here, we shall offer an associated examination from the viewpoint
of the polynomial invariants.

The present investigation demonstrates that definite stability
implications do arise from the intrinsic Riemannian geometry. For
an extreme supergravity configuration, the stabilization problem
finds a natural ground in the framework of the real intrinsic
geometry. We stress that the number of harmonic modes defines
underlying microscopic systems, which in the large charge limit,
allow one to accurately configure the system under the inclusion
of certain fluctuations. Whereas, the respective classical moduli
systems may exactly be defined in terms of a set of scalar fields.
In particular, it turns out that the fluctuation of the scalar
potential, when it is considered as a function of the scalar
fields, characterizes the chosen macroscopic attractor
configuration. In general, we find that the underlying
configuration, which possesses a prime importance towards the
present interest, may be described by the $D$-term potential,
\begin{eqnarray}
V_D(\varphi_i) = (\sum_{i \in \Lambda} \ q_i \ \varphi_i^{\dag}
\varphi_i + b)^2
\end{eqnarray}
Here, the Fayet parameter $b$ divulges the basic nature of the
underlying supersymmetric theory and entails the decay property in
the moduli configurations. At this juncture, it is worth
mentioning that we wish to analyze the walls of stability of the
BPS-configurations, and put our emphasis from the perspective of
the intrinsic real Riemannian geometry. Henceforth, we shall
consider $|q|= 1$. Such a choice follows from the fact that the
moduli fields may be adjusted by an appropriate respective
dilatation. Notice further that the marginal stability requires a
set of uniformly distributed charges and thus the sign should be
taken to be the same as that of the Fayet parameter. Whilst, the
threshold stability requires an alternating set of charges. In
this case, the relative sign of the vacuum value of constituent
scalars is such that the effective potential vanishes for (i)
equal values of the scalars and (ii) $b=0$. This property follows
from the fact that the effective potential is required to ensure
the supersymmetry constraints at the vacuum. Herewith, we shall
focus our attention on the moduli configurations, which are
described by the single and two complex scalar fields. For such
configurations, we shall analyze the walls of BPS stabilities from
the perspective of the real intrinsic geometry.

The physical moduli may thus be described as an intrinsic
fluctuating configuration, whose pair correlations satisfy a set
of interesting planar and hyper-planar stability constraints. In
order to appreciate a formal intrinsic picture, we shall consider
a set of real scalar fields $\{ x_i \}$ with an embedding $V:\{
x_i \} \rightarrow R$, such that there exists a definite scalar
potential for the real scalars. We shall intrinsically analyze the
role of the possible limits: $b \mapsto -b$, $b \mapsto 0$ and $b
\mapsto 1/b$, and thereby offer the intrinsic geometric criterion
to the quantities. In particular, we shall exhibit how the planar
and hyper-planar stability conditions vary under the reflection,
translation and inversion of the Fayet parameter $b$. Notice that
the reflection symmetry of the Fayet term is required, due to the
supersymmetry constraints, and in fact only reflection symmetric
combinations contribute to the physical nature of the underlying
vacuum moduli configuration. In the subsequent section, a similar
role of dilatation and restriction of the Fayet parameter is
offered, from the perspective of the polynomial invariance.

An indispensable relevance of the concepts being divulged in this
paper may thus be structured as a certain intriguing intrinsic
real Riemannian geometry which arises from the nature of
underlying marginal and threshold stable configurations.
Furthermore, it is interesting to note that the marginal stable
configurations allow for BPS to non-BPS decays, whereas the
corresponding threshold stable moduli configurations do not.
Furthermore, we observe that the transformation $b \mapsto -b$
implies that the D-term threshold potentials satisfy
$V_D^{2n}(x_i)= 0$, for all possible physical values of the Fayet
parameter. In particular, it is worth to mention, for a positive
value of the Fayet parameter and two real scalars $\{x_1,x_2\}$,
that the threshold stable configurations require a preferred
surface of constraint, e.g. $<x_1>_0= 0$, $< x_2 >_0= b$, whereas
the choice $< x_1 >_0= b$,$ < x_2 >_0= 0$ leads to a negative
value. In fact, the above cases only offer an illustration of the
consideration, and indeed there exist a whole moduli of surface.
Subsequently, we show, in the subsections $3.2$ and $4.2$, that
similar notions may further be generalized for a larger number of
moduli.

Interestingly, the vanishing value of the Fayet parameter comes up
with an appropriate choice which arrives with the equal vacuum
expectation value $< x_1 >_0=< x_2 >_0$. Thus, we see that the
walls of the marginal and threshold stabilities may be defined by
the limiting moduli configurations which possess no Fayet term. In
this case, one finds that the limit $b = 0$ provides the possible
curve of the moduli stability. The situation under the present
consideration may thus be described via the technology of the
intrinsic Riemannian geometry, where the planar and hyper-planar
stabilities are encoded in the principle minors of the
corresponding covariant metric tensor. As per the comparative
results of the present analysis, we find that a further
application may be explored for the moduli configurations of the
experimental interests, e.g. the phenomenology of string theory.
In fact, there exists a list of such moduli configurations, e.g.,
Higgs moduli, Calabi-Yau moduli, torus moduli, instanton moduli,
topological string moduli, and the moduli pertaining to $D$ and
$M$-particles. These examinations are however left for a separate
investigation.

The rest of the paper has been organized as follows. In section
$2$, we provide a brief review of some of the needful concepts
pertaining to the $D$-term potential, relevance of the Fayet
models, walls of stability and their certain implications arising
from the basics of the real intrinsic Riemannian geometry. In
section $3$, we analyze the moduli stabilization problem for the
marginally stable BPS configurations and thereby explicate the
nature of the stability under the transformation of the Fayet
parameter. In section $4$, we extend the consideration for the
threshold configurations and concentrate on the examination of the
underlying moduli configuration. In sections $5$, we present our
conclusion, and a set of perspective remarks for the future
investigations.
\section{D-term Potentials and Real Intrinsic Geometry}
We begin by considering a brief review of the Fayet models and
related concepts, which would be used in the later sections. In
this section, we shall concisely provide an account for the D-term
potential and stability criteria arising from the real intrinsic
Riemannian geometry. The general details of the Fayet models may
be found in \cite{4a,4b} and the intrinsic geometry has been
explained in \cite{20a,20b,20c,20d}. Some other possible
developments and recent interesting applications of the $D$-term
and $F$-term potentials may be found in Ref.\cite{21}. What
follows next is that one may consider a set of complex scalar
fields $\{ \varphi_i\}_{i=1}^n$ with the respective
$U(1)$-charges$\{ q_i\}_{i=1}^n$. Then, the leading order D-term
potential, in an appropriate normalization convention, may be
expressed as,
\begin{eqnarray}
V_D(\varphi_i) = (\sum_{i \in \Lambda_0} \ \varphi_i^{\dag}
\varphi_i + b)^2
\end{eqnarray}
The realization of the scalar fields may be accomplished by
defining $\varphi_i = x_i + ix_{i+1}$, and thus the concerned
D-term potential reduces to
\begin{eqnarray}
V_D(x_i) = (\sum_{i \in \Lambda} \ x_i^2 + b)^2
\end{eqnarray}
As mentioned in the introduction, the issue under consideration
may be analyzed by extremizing the D-term potential $V_D(x_i)$.
The set of critical points $\mathcal{C}:= \{ x_i^0|\
i=1,2,...,2n\}$ of the potential can be easily determined by the
condition $\partial V_D(x_i) = 0$. While the stability of the
underlying configuration may straightforwardly be achieved by
demanding the positivity of the Hessian matrix of $V_D(x_i)$.
Thence, an arbitrary configuration is stabilized according to the
 following definition
\begin{eqnarray}
\partial_i \partial_j V_D(x_i)\vert_{x_i=x_i^0} = g_{ij}
\end{eqnarray}
We may however easily notice, under the consideration of extreme
moduli space geometry [17], that the ordinary Hessian matrix
$\partial_i \partial_j V_D(x_i)$ defines a symmetric bilinear
form, and thus supplies a real intrinsic metric tensor $g_{ij}$.
Thus, the stability analysis of the concerned configuration may be
performed, in terms of the positivity of the principle minors of
the covariant metric tensor $g_{ij}$, see for detailed physical
applications of the intrinsic geometry
\cite{12a,12b,12c,12d,12e,12f,12g,12h,12i,12j,12k,12l,12m,12n,
12o,12oo,12ooo,12o1,12o2,12p,12q,12r,12s,12t,12u,20a,20b,20c,20d,21}.

The linear stability may simply be obtained by demanding the
positivity of the principle components of the real metric tensor.
Thus, the system is stable along an intended dimension $n$, if the
respective component satisfies $g_{ij}>0$. Furthermore, the
configuration is stable on the two dimensional surfaces being
defined by the coordinate chart $\{x_1, x_2\}$, if the concerned
determinant of the metric tensor satisfies,
\begin{eqnarray}
g_2 = g_{11}g_{22}- g_{12}^2 > 0
\end{eqnarray}
Moreover, the chosen solution remains stable on the three
dimensional hyper-surface, if the determinant of the metric tensor
satisfies,
\begin{eqnarray}
g_3 = g_{11}(g_{22}g_{33}- g_{23}^2)- g_{12}(g_{12}g_{33}-
g_{13}g_{23})+ g_{13}(g_{12}g_{23}- g_{13}g_{22})> 0
\end{eqnarray}
Similarly, an arbitrary system of moduli turns out to be stable,
as the $m \le 2n$-dimensional hyper-surface, if the concerned
principle minors and the determinant of the metric tensor remain
positive definite quantities. In this viewpoint, the full
configuration is said to be stable against the simultaneous
fluctuations of the moduli fields, if the underlying determinant
of the metric tensor remains a positive definite quantity, over a
range of interest of the parameters.

In the above definition, a moduli configuration is said to be
completely stable, if the set \begin{eqnarray}\mathcal{B}:=
\{g_{ij},g_{i}, g; x_i \in M_{2n}, \forall
i=1,2,...,2n\}\end{eqnarray} remains positive definite. It is
worth mentioning that a moduli is stabilized, if an arbitrary
scalar $x_i \in M_{2n}$ also satisfies the same requirement, i.e.
it is an element of the set $\mathcal{B}$.

A set of scalars $\{x_i \in M_{2n} \}$ are said to be (hyper)
correlated, if the components of the underlying Riemann covariant
tensor $R_{ijkl}$ remain non-vanishing for some given indices $i,
j, k, l$. In particular, the scalar curvature signifies an average
correlation volume for the constituent moduli field configuration.
The present paper concentrates on the decaying BPS configurations,
which are associated with the $D$-term potentials. Subsequently,
we shall focus our attention on the marginal and threshold walls
of stabilities.

From the perspective of the $D$-term potential, there exists no
set of critical points $\{x_1^0, x_2^0 \in \mathcal{C} \}$, such
that the corresponding potential $V (x_1^0, x_2^0)= 0$. In this
case, if there are some decays allowed, then the moduli
configuration thus obtained must be a non-BPS system. A similar
analysis is straightforward for the multi-complex scalar
configurations, as well. In particular, we may note that there
does not exist a set of vacuum scalars $\{x_1^0, x_2^0,...,
x_n^0\}$ such that the  $V (x_i^0)= 0$. Consequently the daughter
system must be a non-BPS configuration. Notice further that the
present analysis is not limited to a single final configuration.
In fact, the investigation in question may be easily carried
forward for an arbitrary union of the daughter configurations, and
thus for the multicentered solutions.

A possible extension of the present investigation may be
accomplished, for the general non-extreme configuration, by
defining the Hessian of the D-term potential as, $g_{ij}=
D_iD_jVD(x_i)$, see for details
\cite{13a,13b,13c,13d,13e,13f,14,15,16}. Although, the underlying
physical interpretations may not remain quite the same as those of
the extreme configurations, it may however be noted that the
scalar fields defining the underlying vacuum manifold may not be
globally stabilized. It is hence natural to extend an
understanding of the attractor flows, what we have thus studied,
in the context of the simplest Fayet models, defined by the
leading order D-term configurations. Nevertheless, the leading
order potential function is particularly suited for the present
analysis, defining an intrinsic quadratic form which assumes no
supersymmetry. Thus, the present analysis considerably simplifies
the underlying geometric computations and the possible
investigation of the transformations concerning the Fayet
parameter.

At this point, it is worth to note that our geometrical analysis
may equally be applied to the other black brane solutions, which
possess a definite moduli space configuration. Interestingly, the
various extensions of the intrinsic geometry may further be
analyzed apart from the extreme configurations. In fact, it may be
noted further that the underlying investigations find a set of
intriguing realizations, from the perspective of the wall crossing
phenomena, for the possible values of the parameters of underlying
moduli space configurations. Apart from the implications following
from the leading order $D$-term potential, there exists a wide
class of effective theories with and without the cosmological
constant, extremal as well as non extremal black brane solutions,
that might be further explored under the agenda of the present
consideration.

The moduli space geometry has in turn given very crucial insights
to the geometric understanding of a class of higher derivative
corrected extremal black hole solutions. It may be instructive to
pursue geometric and algebraic dispositions associated with
various black hole potentials, with an inclusion of arbitrary
higher derivative curvature terms, for various space-time
dimensions. In particular, the underlying analysis may be
investigated for the attractor stability of $AdS_2 \times S^{D-2}$
near horizon geometry, for an arbitrary extremal black hole.
Importantly, the case of the Calabi-Yau black holes may
investigated under the string duality transformations, containing
certain monodromy invariant parameters. Thus, one may
intrinsically analyze the phenomenon of the wall crossing, from
the perspective of the Calabi-Yau compactification.

What follows next is that the nature of the supersymmetric moduli
may be characterized by the charges carried by the configuration.
Thus, depending on the sign of the underlying charges, we shall
analyze the intrinsic geometric issues and thereby describe the
nature of the wall crossing phenomena, for the decaying single
charge and two charge moduli configurations. In the next section,
we present the case of the marginal wall of stability and possible
insights arising from the real intrinsic geometry.

\section{Marginal Stability}
In this section, we focus our attention on uniform charge
distribution for the scalar fields $\varphi_i$, such that each of
them has an equal unit charge. In other words, we shall consider
that the charges carried by each of the scalar fields are $q_i =
1, \forall i$. To make the presentation lucid, we shall analyze
the associated moduli configurations for the case of one and two
$U(1)$ scalars, which respectively involve two and four real
scalar fields. Sequentially, it turns out further that the general
consideration may similarly be illustrated, for an arbitrary
number of abelian scalar fields.
\subsection{Single Complex Scalar}
In order to offer a flavor of the present analysis, let us
consider a single complex scalar field $\varphi_i:= (x_1, x_2) \in
U(1)$. Thus, with an appropriate normalization, the potential of
interest may be expressed as,
\begin{eqnarray}
V(x_1,x_2) := (x_1^2+x_2^2+b)^2
\end{eqnarray}
A straightforward computation shows that the components of the
metric tensor satisfy a set of quadratic polynomials,
\begin{eqnarray}
g_{x_ix_i}&=& 12 x_i^2+4 x_j^2+4 b, i,j=1,2; i \neq j \nonumber \\
g_{x_1x_2}&=& 8 x_1x_2 
\end{eqnarray}
We thus see that the principle components of the intrinsic metric
tensor are symmetric quadratic polynomials in the real scalars,
whereas the off-diagonal components are symmetric quadratic
monomials. From the definition of the intrinsic geometry, we see
that the determinant of the metric tensor takes a well-defined
positive-definite quadratic polynomial form,
\begin{eqnarray}
g = 16 (b^2+4( x_1^2+x_2^2) b+3(x_1^2+ x_2^2)^2)
\end{eqnarray}
Thence, we observe that the determinant of the metric tensor is a
quadratic polynomial in the Fayet parameter defining the D- term
potential of the underlying supersymmetry breaking configuration.
Furthermore, we may easily compute the Christoffel connections,
Riemann covariant tensors, and the associated Ricci tensors, as
well. for a given intrinsic surface. Herewith, we find that the
scalar curvature takes the following quotient expression,
\begin{eqnarray}
R = -b \frac{(x_1^2+x_2^2)}{(b^2+4( x_1^2+x_2^2) b+3(x_1^2+
x_2^2)^2)^2}
\end{eqnarray}
In order to understand the implication of the Fayet parameter, we
shall now consider the various possible physically interesting
limiting cases, and thereby focus our attention on the underlying
transformations. For example, we see, in the limit $b= 0$, that
the determinant of the metric tensor reduces to a positive
expression
\begin{eqnarray}
g = 48(x_1^2+x_2^2)^2
\end{eqnarray}
which is a positive definite function for all values of the moduli
fields $x_1,x_2$. Whilst, it follows that the corresponding scalar
curvature vanishes identically. This implies that the underlying
system becomes a non-interesting statistical system for $b=0$.

Furthermore, under the reflection $b= -a$, we may easily notice
that the determinant of the metric tensor satisfies
\begin{eqnarray}
g = 16 (a^2-4(x_1^2+ x_2^2) a+ 3(x_1^2+ x_2^2)^2
\end{eqnarray}
At such a negative value of the Fayet parameter, $b=-a$, the
scalar curvature transforms to
\begin{eqnarray}
R= a \frac{(x_1^2+x_2^2)}{(a^2-4(x_1^2+ x_2^2) a+ 3(x_1^2+
x_2^2)^2}
\end{eqnarray}
This shows that the marginal configuration, as per the definition
of the $D$-term potential, remains statistically interacting,
under the reflection of the Fayet parameter.

Notice that similar conclusions are obtained under the inversion
$b = 1/c$. In particular, one finds that the determinant of the
metric tensor reads
\begin{eqnarray}
g = \frac{16}{c^2} (c x_1^2+c x_2^2+1) (3 c x_1^2+3 c x_2^2+1)
\end{eqnarray}
Correspondingly, we see that the scalar curvature reduces to
\begin{eqnarray}
R = - c^3 \frac{(x_1^2+x_2^2)}{(c x_1^2+c x_2^2+1)^2(3 c x_1^2+3 c
x_2^2+1)^2}
\end{eqnarray}

\begin{figure}
\hspace*{0.5cm}
\includegraphics[width=8.0cm,angle=-90]{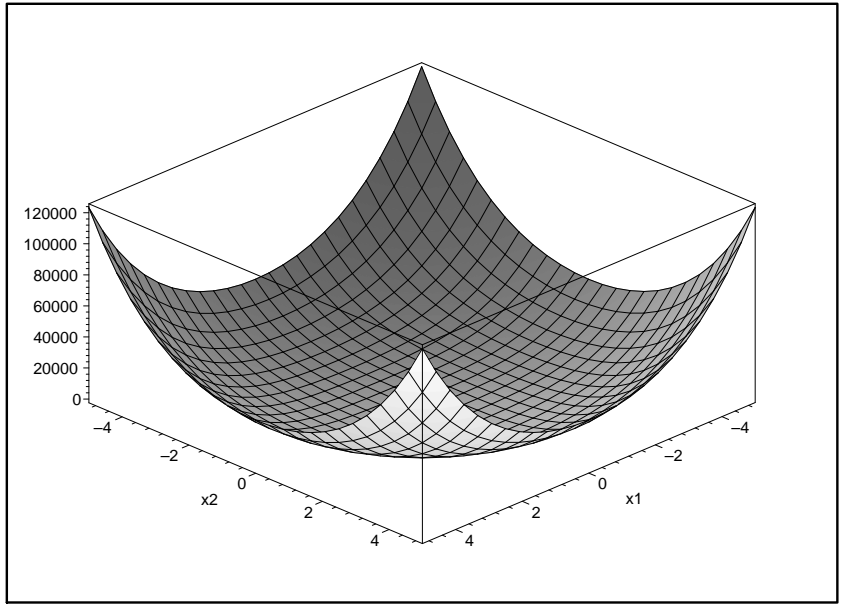}
\caption{The determinant of the metric tensor plotted as a
function of the moduli fields, $x_1, x_2$, describing the
fluctuations in the marginal configuration.} \label{marginal1g}
\vspace*{0.5cm}
\end{figure}

\begin{figure}
\hspace*{0.5cm}
\includegraphics[width=8.0cm,angle=-90]{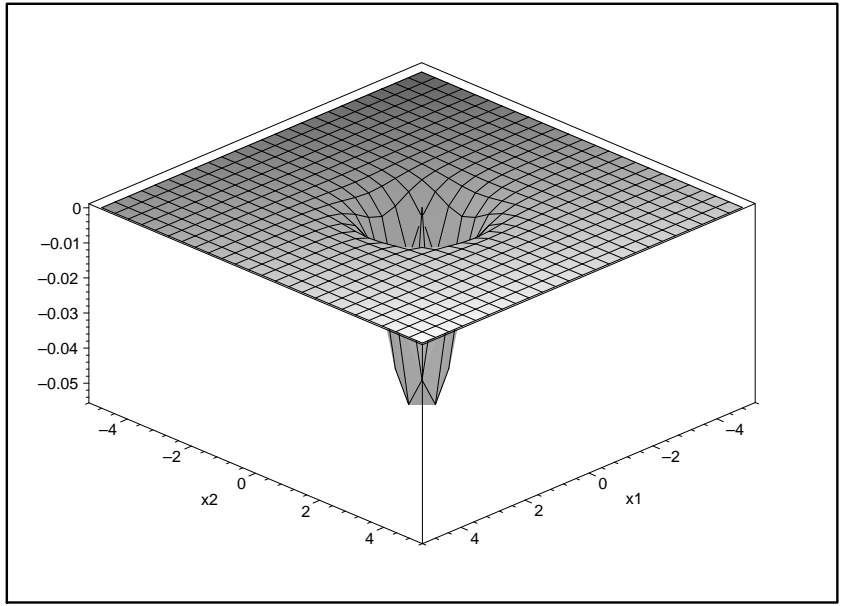}
\caption{The scalar curvature plotted as a function of the moduli
fields, $x_1, x_2$, describing the fluctuations in the marginal
configuration.} \label{marginal1R} \vspace*{0.5cm}
\end{figure}

Finally, it is obvious for equal values of the scalar fields, $x_1
= x$ and $x_2 = x $, that the components of the metric tensor take
the following values
\begin{eqnarray}
g_{x_ix_i}&=& 16 x^2+4 b \nonumber \\
g_{x_1x_2}&=& 8 x^2 
\end{eqnarray}
The corresponding determinant of the metric tensor simplify to the
following quadratic polynomial $g = 192x^4+128bx^2+ 16b^2$ in $b$.
In this limit, it is also not difficult to see that the scalar
curvature reduces to the following expression
\begin{eqnarray}
R = -2 b\frac{ x^2}{(b^2+8 x^2 b+12 x^4)^2}
\end{eqnarray}
Interestingly, we find that the sign of the moduli interactions
depends only on the sign of the Fayet parameter. Specifically, the
underlying system is attractive, for a positive value of the Fayet
parameter. The graphical notions of the stability and global
moduli interactions are respectively depicted in the
Figs(\ref{marginal1g}, \ref{marginal1R}). Over a range of moduli
fields $\{x_1, x_2\}$, the figures henceforth have been plotted
for the choice of a unit Fayet parameter $b=1$.
\subsection{Multiple Complex Scalars}
We shall now provide a general flavor of the analysis presented in
the foregoing subsection. This is accomplished by taking a set of
complex scalar fields $\{ \varphi_i| \forall i \in \Lambda,
\varphi_i \in U(1) \}$, where the index $\Lambda$ is taken to be
the set of a finite cardinality. First of all, it is natural to
concentrate on the case of the two complex scalar fields $ \{
\varphi_1, \varphi_2 \}$, thus applying the previously defined
analysis for the four real scalars $(x_1, x_2, x_3, x_4)$.
Nevertheless, it may be further envisaged that it is possible to
analyze the nature of the marginal stability for the general
moduli configurations with an arbitrary $\Lambda$, as well. Let us
focus our attention on the marginal stability of the leading order
configurations. In particular, let us consider the D-term
potential as a function of the four real scalar fields, viz.,
$x_a= (x_1,x_2,x_3,x_4)$. Thus, it is immediate to see that the
underlying potential may be expressed as,
\begin{eqnarray}
V(x_1,x_2,x_3,x_4) := (x_1^2+x_2^2+x_3^2+x_4^2+b)^2
\end{eqnarray}
A straightforward calculation implies that the covariant
components of the metric tensor are,
\begin{eqnarray}
g_{x_ix_i}&=& 12 x_i^2+4 \sum_{j \neq i }x_j^2 +4 b, \ i= 1,2,3,4 \nonumber \\
g_{x_ix_j} &=& 8 x_ix_j, \ \forall i \neq j, \ i,j= 1,2,3,4
\end{eqnarray}
In this case, we observe further that the principle components of
the intrinsic metric tensor are quadratic polynomials in the real
scalars, while the diagonal components are symmetric monomials.
Apart from the number of polynomials, this property remains
exactly the same, as mentioned in the foregoing subsection. In
order to analyze the planar and hyper planar stability conditions,
we now need to consider the principle minors of the metric tensor.
Thence, we find that the planar principle minor, defined as $g_2:=
g_{11}g_{22}-g_{22}^2$, leads to the following quadratic
polynomial,
\begin{eqnarray}
g_2:=  g_{22}b^2+ g_{21}b+ g_{20}
\end{eqnarray}
where the coefficients of the above equation are given by the
following expressions
\begin{eqnarray}
g_{22}&:=&16  \nonumber \\
g_{21}&:=&(32 x_3^2+32 x_4^2+64 x_1^2+64 x_2^2) \nonumber \\
g_{20}&:=&64 x_2^2 x_3^2+64 x_2^2 x_4^2+96 x_2^2 x_1^2+48 x_2^4+16
x_3^4 \nonumber \\ && +16 x_4^4+48 x_1^4+64 x_1^2 x_4^2+64 x_1^2
x_3^2+32 x_3^2 x_4^2
\end{eqnarray}
Whilst, the hyper-planar stability may be examined by the third
principle minor, defined as $g_3 = g_{11}(g_{22}g_{33}- g_{23}^2)-
g_{12}(g_{12}g_{33}- g_{13}g_{23})+ g_{13}(g_{12}g_{23}-
g_{13}g_{22})$. It is not difficult to show that $g_3$ simplifies
to a cubic polynomial
\begin{eqnarray}
 g_3 := g_{33}b^3+ g_{32} b^2+ g_{31}b +g_{30}
\end{eqnarray}
where the coefficients of the above equation take the following
expressions
\begin{eqnarray}
g_{33}&:=& 64 \nonumber \\
g_{32}&:=&192 x_4^2+320 x_1^2+320 x_2^2+320 x_3^2 \nonumber \\
g_{31}&:=&448 x_3^4+384 x_1^2 x_4^2+192 x_4^4+1152 x_1^2 x_3^2+640
x_2^2 x_4^2 \nonumber \\ &&
+896 x_2^2 x_1^2+448 x_2^4+896 x_2^2 x_3^2+448 x_1^4+640 x_3^2 x_4^2 \nonumber \\
g_{30}&:=& 896 x_1^2 x_2^2 x_3^2+576 x_1^2 x_2^4+832 x_1^2
x_3^4+64 x_1^2 x_4^4+832 x_1^4 x_3^2 \nonumber \\ && +192 x_1^4
x_4^2+576 x_1^4 x_2^2+576 x_2^4 x_3^2+448 x_2^4 x_4^2+448 x_3^4
x_4^2 \nonumber \\ && +576 x_2^2 x_3^4+320 x_2^2 x_4^4+192
x_3^6+64 x_4^6+1024 x_2^2 x_1^2 x_4 x_3 \nonumber \\ && +192
x_1^6+192 x_2^6+320 x_3^2 x_4^4+128 x_1^2 x_2^2 x_4^2+896 x_2^2
x_3^2 x_4^2 \nonumber \\ && +896 x_1^2 x_3^2 x_4^2
\end{eqnarray}
For arbitrary values of the scalar fields and the Fayet parameter,
it is not difficult to see that the determinant of the metric
tensor is given by,
\begin{eqnarray}
 g := g_{44} b^4+ g_{43}b^3+g_{42} b^2+g_{41} b+g_{40}
\end{eqnarray}
where the coefficients of the above equation can be expressed as
follows
\begin{eqnarray}
g_{44}&:=& 256  \nonumber \\
g_{43}&:=& 1536 x_3^2+1536 x_2^2+1536 x_4^2+1536 x_1^2  \nonumber \\
g_{42}&:=& 3072 x_3^4+6144 x_1^2 x_3^2+6144 x_1^2 x_4^2+3072
x_2^4+6144 x_2^2 x_3^2  \nonumber \\ &&
+3072 x_1^4+6144 x_2^2 x_1^2+6144 x_3^2 x_4^2+3072 x_4^4+6144 x_2^2 x_4^2  \nonumber \\
g_{41}&:=& 7680 x_2^4 x_4^2+2560 x_3^6+7680 x_1^4 x_3^2+7680 x_2^4
x_3^2+2560 x_2^6  \nonumber \\ && +7680 x_1^2 x_2^4+7680 x_3^4
x_4^2+7680 x_1^4 x_4^2+7680 x_2^2 x_3^4+15360 x_1^2 x_3^2 x_4^2
\nonumber \\ && +15360 x_1^2 x_2^2 x_3^2+7680 x_1^2 x_3^4+7680
x_1^2 x_4^4+2560 x_4^6+7680 x_3^2 x_4^4  \nonumber \\ &&
+2560 x_1^6+7680 x_2^2 x_4^4+15360 x_1^2 x_2^2 x_4^2+7680 x_1^4 x_2^2+15360 x_2^2 x_3^2 x_4^2  \nonumber \\
g_{40}&:=& 3072 x_2^6 x_3^2+4608 x_2^4 x_3^4+4608 x_2^4 x_4^4+3072
x_2^6 x_4^2+3072 x_2^2 x_3^6  \nonumber \\ && +3072 x_2^2
x_4^6+4608 x_1^4 x_2^4+3072 x_1^2 x_2^6+3072 x_1^6 x_2^2+3072
x_4^6 x_3^2  \nonumber \\ && +4608 x_3^4 x_4^4+3072 x_3^6
x_4^2+4608 x_1^4 x_3^4+3072 x_1^2 x_3^6+3072 x_1^2 x_4^6 \nonumber
\\ && +4608 x_1^4 x_4^4+3072 x_1^6 x_3^2+3072 x_1^6
x_4^2+9216 x_1^2 x_3^4 x_4^2+9216 x_1^4 x_2^2 x_3^2  \nonumber \\
&& +9216 x_1^2 x_2^4 x_3^2+9216 x_1^2 x_2^2 x_3^4+9216 x_1^2 x_4^4
x_2^2+9216 x_1^2 x_4^4 x_3^2  \nonumber \\ && +9216 x_1^4 x_3^2
x_4^2+9216 x_1^4 x_4^2 x_2^2+9216 x_2^4 x_3^2 x_4^2+9216 x_2^4
x_4^2 x_1^2  \nonumber \\ && +9216 x_2^2 x_3^4 x_4^2+9216 x_2^2
x_4^4 x_3^2+768 x_1^8+768 x_2^8+768 x_4^8  \nonumber \\ && +768
x_3^8+18432 x_2^2 x_1^2 x_4^2 x_3^2
\end{eqnarray}
An analogous analysis may as well be performed easily for the
concerned global intrinsic geometric invariant quantities. In
particular, it is not difficult to see that the underlying Ricci
scalar of the fluctuations is,
\begin{eqnarray} R = -\frac{9}{2}
\frac{\sum_{i=1}^4x_i^2}{(b+3 \sum_{j=1}^4 x_j^2)} \frac{1}{(b^2+
r_{m1} b+ r_{m0})}
\end{eqnarray}
where the polynomial functions $r_{m1}$ and $r_{m0}$ are defined
as follows
\begin{eqnarray}
r_{m1}&:=&4 (\sum_{i=1}^4 x_i^2)  \nonumber \\
r_{m0}&:=&3 (\sum_{i=1}^4 x_i^2)^2
\end{eqnarray}

\begin{figure}
\hspace*{0.5cm}
\includegraphics[width=8.0cm,angle=-90]{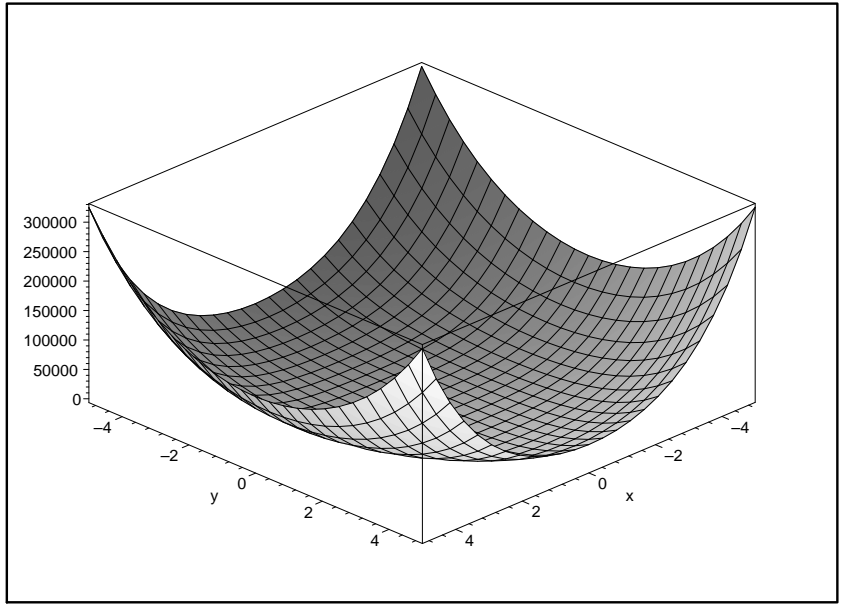}
\caption{The surface minor plotted as a function of the moduli
fields, $x, y$, describing the fluctuations in the marginal
configuration.} \label{marginal2g2} \vspace*{0.5cm}
\end{figure}

\begin{figure}
\hspace*{0.5cm}
\includegraphics[width=8.0cm,angle=-90]{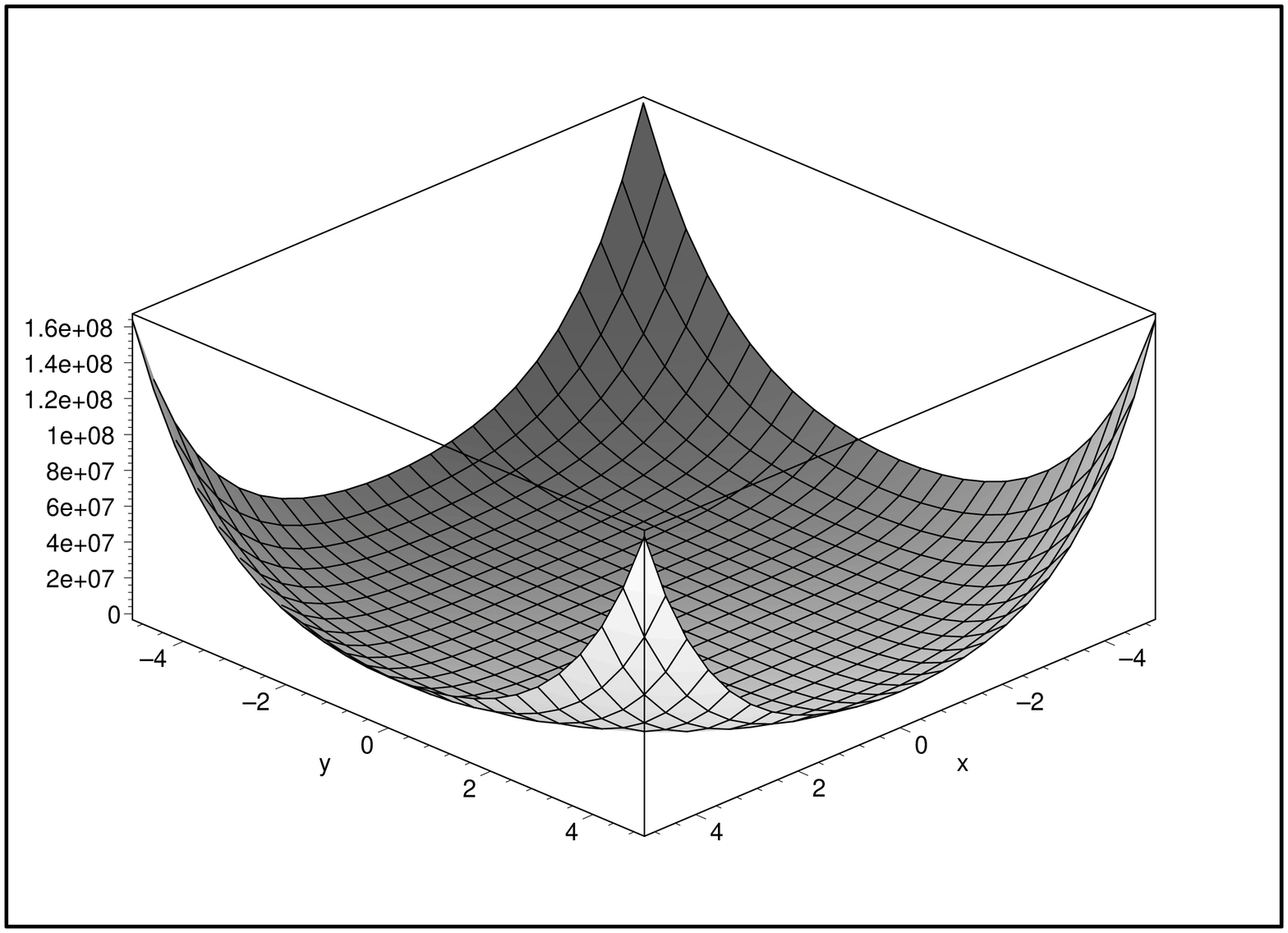}
\caption{The hypersurface minor plotted as a function of the
moduli fields, $x, y$, describing the fluctuations in the marginal
configuration.} \label{marginal2g3} \vspace*{0.5cm}
\end{figure}

\begin{figure}
\hspace*{0.5cm}
\includegraphics[width=8.0cm,angle=-90]{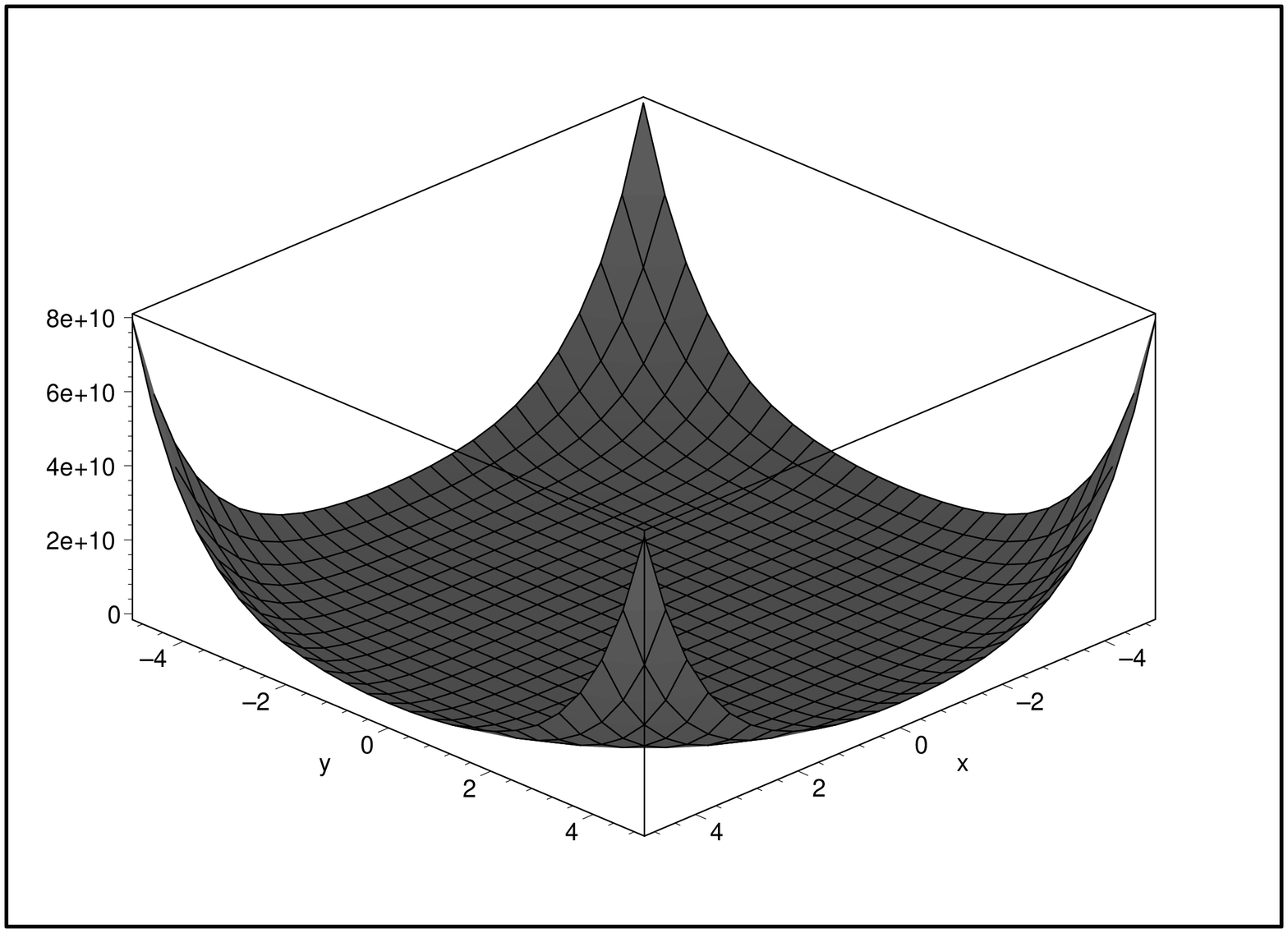}
\caption{The determinant of the metric tensor plotted as a
function of the moduli fields, $x, y$, describing the fluctuations
in the marginal configuration.} \label{marginal2g} \vspace*{0.5cm}
\end{figure}

\begin{figure}
\hspace*{0.5cm}
\includegraphics[width=8.0cm,angle=-90]{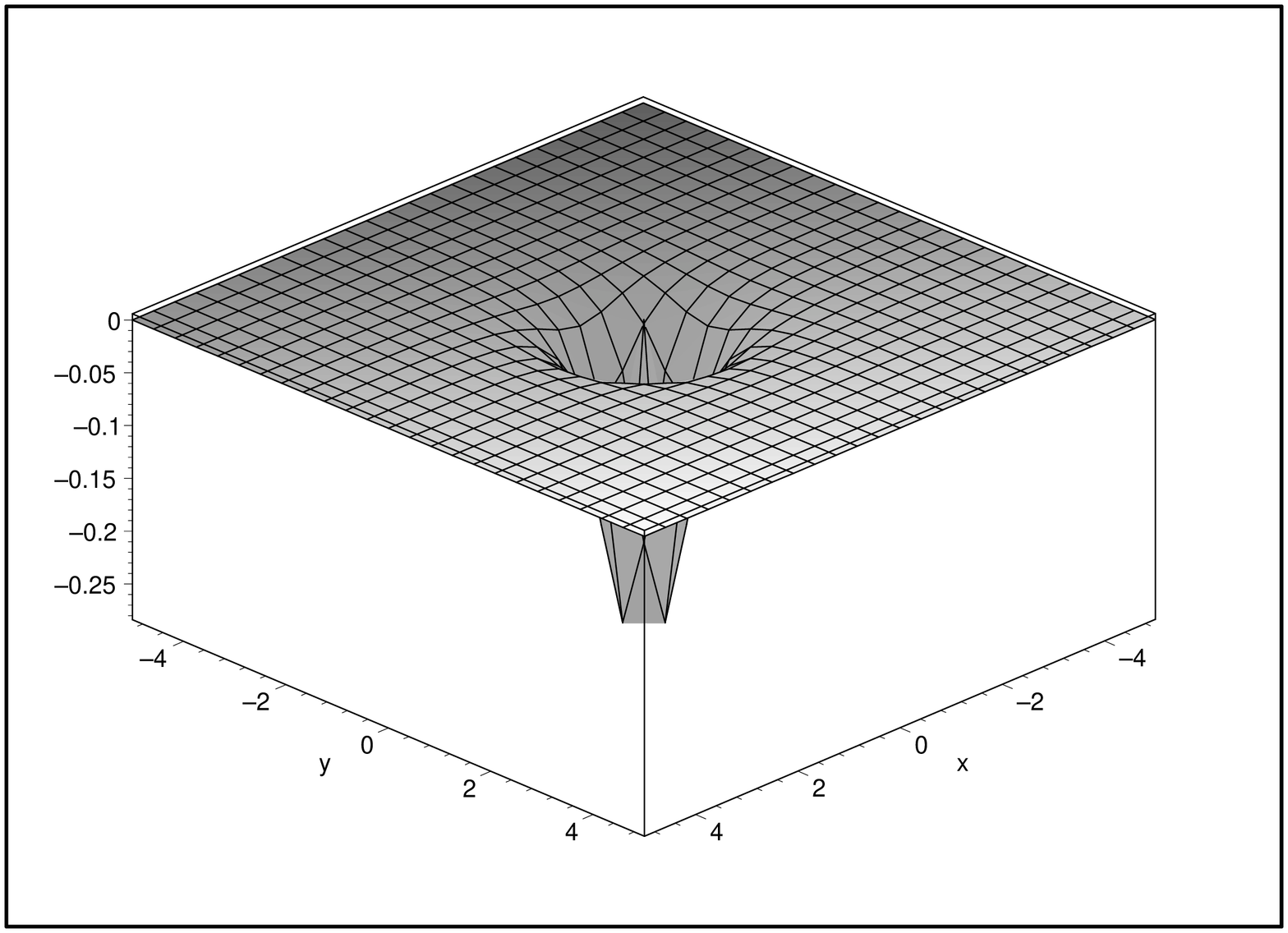}
\caption{The scalar curvature plotted as a function of the moduli
fields, $x, y$, describing the fluctuations in the marginal
configuration.} \label{marginal2R} \vspace*{0.5cm}
\end{figure}

For a pairwise equal value of the moduli fields, viz., $x_1= x_2=
x$, $x_3= x_4= y$ and for a positive value of the Fayet parameter,
e.g., $b=1$, the principle minors are given by
\begin{eqnarray}
g_2 &=& 256 x^2 y^2+192 x^4+16+64 y^2+64 y^4+128 x^2
\end{eqnarray}
\begin{eqnarray}
g_3 &=& 64+640 x^2+512 y^2+3072 x^2 y^2+1792 x^4 \nonumber \\ &&
+1280 y^4+4096 x^4 y^2+3584 x^2 y^4+1024 y^6+1536 x^6
\end{eqnarray}
\begin{eqnarray}
g &=& 256+3072 x^2+3072 y^2+24576 x^2 y^2+12288 x^4+12288 y^4
\nonumber \\ && +49152 x^2 y^6+73728 x^4 y^4+49152 x^6 y^2+61440
x^4 y^2 \nonumber \\ && +61440 x^2 y^4+12288 x^8+12288 y^8+20480
y^6+20480 x^6
\end{eqnarray}
It follows further that the associated scalar curvature is given
by the following formula
\begin{eqnarray}
R = -9 \frac{(x^2+ y^2)}{(1+6 x^2+6 y^2)(1+8 y^2+8 x^2+12 x^4+12
y^4+24 x^2 y^2)}
\end{eqnarray}
The qualitative behavior of the principle minors $g_2, g_3$,
determinant of the metric tensor $g$ and the scalar curvature $R$
has been respectively shown in the following 3 dimensional
figures: Figs.(\ref{marginal2g2}, \ref{marginal2g3},
\ref{marginal2g}, \ref{marginal2R}).

\begin{figure}
\hspace*{0.5cm}
\includegraphics[width=8.0cm,angle=-90]{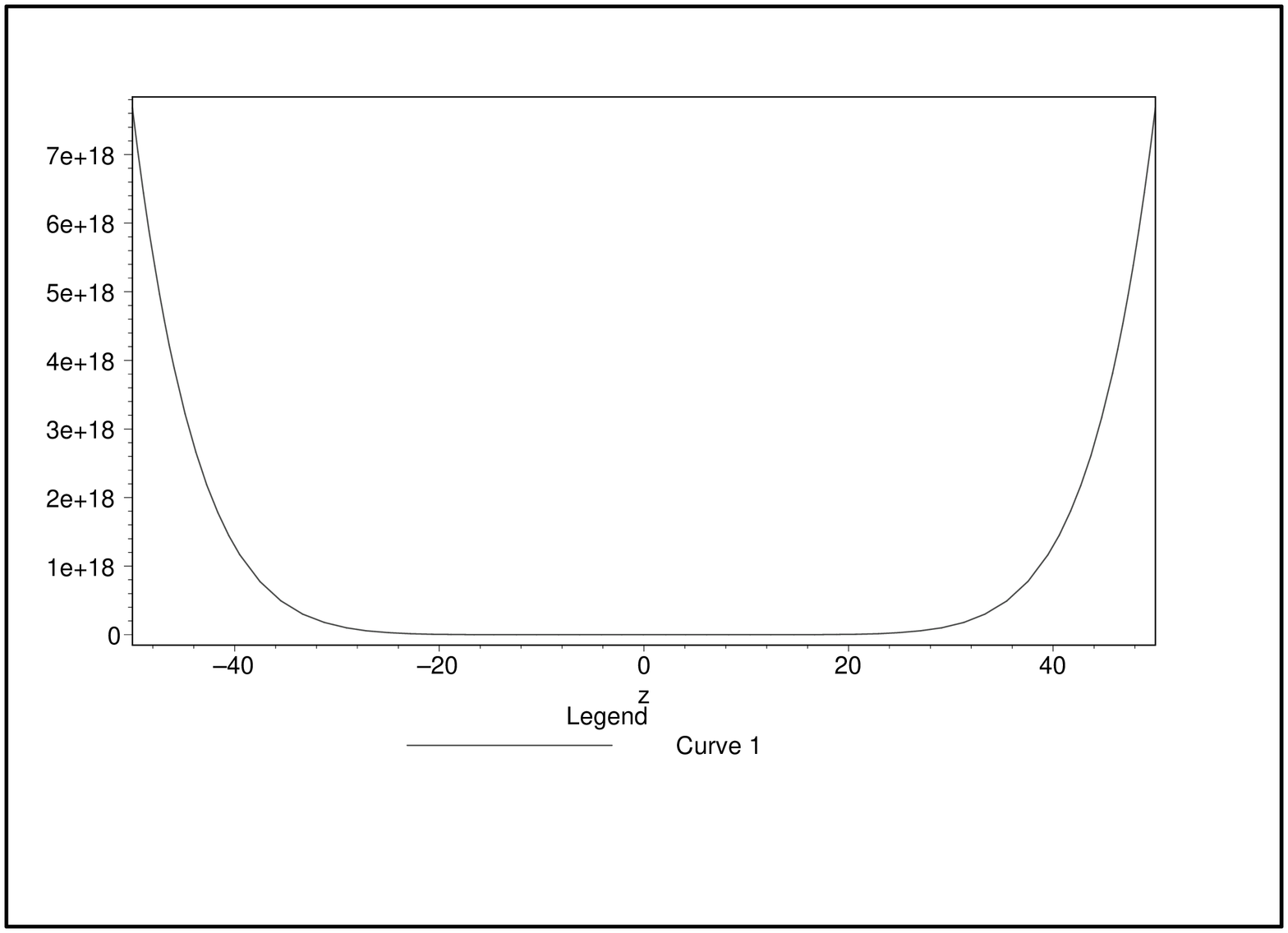}
\caption{The determinant of the metric tensor plotted as a
function of the moduli fields, $z$, describing the fluctuations in
the marginal configuration.} \label{marginal2g2dim}
\vspace*{0.5cm}
\end{figure}

\begin{figure}
\hspace*{0.5cm}
\includegraphics[width=8.0cm,angle=-90]{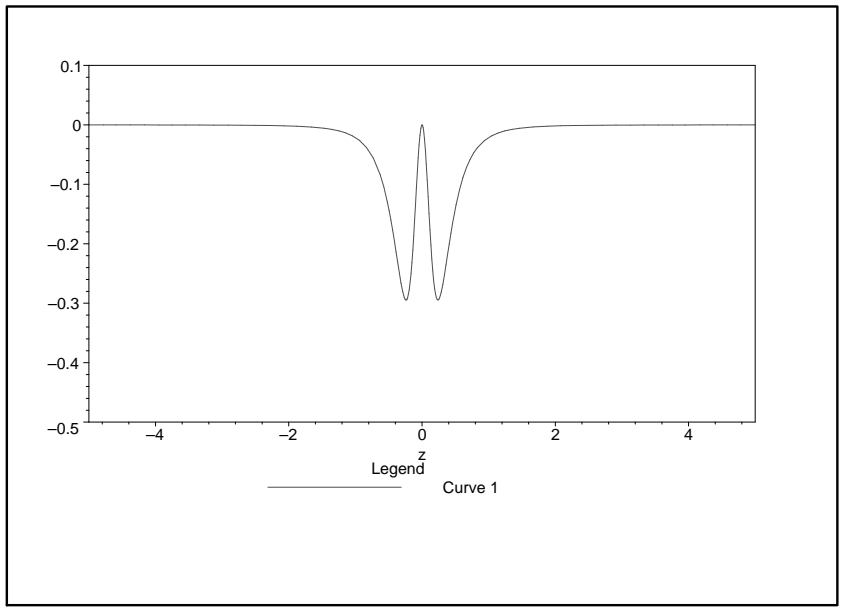}
\caption{The scalar curvature plotted as a function of the moduli
fields, $z$, describing the fluctuations in the marginal
configuration.} \label{marginal2R2dim} \vspace*{0.5cm}
\end{figure}

For equal values of the scalar fields, viz, $x=z$, $y=z$ and
$b=1$, the limiting determinant of the metric tensor possesses the
following qualitative behavior. Specifically, the two dimensional
nature of the determinant of the metric tensor $g$ and the scalar
curvature $R$ has been respectively shown in the following plots:
Figs.(\ref{marginal2g2dim}, \ref{marginal2R2dim}).

For equal values of the scalar fields, $x_i= x$, we find, for any
general Fayet parameter $b$, that the planar stability may be
determined from the polynomial expressions of the corresponding
principle minors, viz., $\{g_2, g_3\}$ and the determinant of the
metric tensor $g$, as the highest principle minor. In this case,
it turns out that the respective limiting values of the principle
minors and determinant of the metric tensor are given by
\begin{eqnarray}
 g_2 := 16 b^2+192 x^2 b+512 x^4
\end{eqnarray}
\begin{eqnarray}
 g_3 := 64 b^3+1152 x^2 b^2+6144 x^4 b+10240 x^6
\end{eqnarray}
\begin{eqnarray}
 g := 256 b^4+6144 x^2 b^3+49152 x^4 b^2+163840 x^6 b+196608 x^8
\end{eqnarray}
The associated scalar curvature is given by the
\begin{eqnarray}
 R = -18 \frac{x^2}{(12 x^2+b)(b^2+16 x^2 b+48 x^4)}
\end{eqnarray}

It may thus be noted that the restricted two dimensional moduli
configuration has the same degree of the polynomial in the Fayet
parameter, as that of the moduli configuration with a single
complex scalar field. Nevertheless, the stability of the two
moduli configuration requires the positivity of the hyper-planar
minor, viz., $g_3$ and that of the other higher principle minors,
e.g. determinant of the metric tensor $g$.

Such stability constraints may further be understood from the fact
that the three dimensional hyper-surface satisfies a third degree
polynomial equation in the Fayet parameter $b$.

For the case of the limiting Fayet parameter $b=0$, we herewith
observe that the local moduli pair correlations reduce to the
following expressions
\begin{eqnarray}
 g_{x_ix_i} &=& 12 x_i^2+4 \sum_{j \neq i} x_j^2 \nonumber \\
 g_{x_ix_j} &=& 8 x_i x_j, i \neq j, i,j=1,2,3,4
\end{eqnarray}
In this case, we find that the surface minor take the following
pure quartic form
\begin{eqnarray}
g_2 &=& 64 x_2^2 x_3^2+64 x_2^2 x_4^2+96 x_2^2 x_1^2+48 x_2^4+16
x_3^4 \nonumber \\&& +16 x_4^4+48 x_1^4+64 x_1^2 x_4^2+64 x_1^2
x_3^2+32 x_3^2 x_4^2
\end{eqnarray}
For the given moduli fields $\{x_1,x_2,x_3,x_4\}$, the
hypersurface minor is given by the following four variable
homogeneous degree 6 polynomial
\begin{eqnarray}
g_3 &=& 896 x_1^2 x_2^2 x_3^2+576 x_1^2 x_2^4+832 x_1^2 x_3^4+64
x_1^2 x_4^4+832 x_1^4 x_3^2 \nonumber \\&& +192 x_1^4 x_4^2+576
x_1^4 x_2^2+576 x_2^4 x_3^2+448 x_2^4 x_4^2+448 x_3^4 x_4^2
\nonumber \\&& +576 x_2^2 x_3^4+320 x_2^2 x_4^4+192 x_3^6+64
x_4^6+1024 x_2^2 x_1^2 x_4 x_3 \nonumber \\&& +192 x_1^6+192
x_2^6+320 x_3^2 x_4^4+128 x_1^2 x_2^2 x_4^2+896 x_2^2 x_3^2 x_4^2
\nonumber \\&& +896 x_1^2 x_3^2 x_4^2
\end{eqnarray}

For the vanishing Fayet parameter $b = 0$, we may now observe that
the determinant of the metric tensor reduces to the following
homogeneous degree 8 polynomial
\begin{eqnarray}
g &=& 9216 x_2^4 x_4^2 x_1^2+9216 x_1^2 x_4^4 x_2^2+18432 x_2^2
x_1^2 x_4^2 x_3^2+9216 x_1^2 x_2^2 x_3^4 \nonumber \\&& +3072
x_2^6 x_3^2+4608 x_2^4 x_3^4+4608 x_2^4 x_4^4+3072 x_2^6
x_4^2+3072 x_2^2 x_3^6 \nonumber \\&& +3072 x_2^2 x_4^6+4608 x_1^4
x_2^4+3072 x_1^2 x_2^6+3072 x_1^6 x_2^2+3072 x_4^6 x_3^2 \nonumber
\\&& +4608 x_3^4 x_4^4+3072 x_3^6 x_4^2+4608 x_1^4 x_3^4+3072
x_1^2 x_3^6+3072 x_1^2 x_4^6 \nonumber \\&& +9216 x_2^2 x_4^4
x_3^2+9216 x_1^4 x_4^2 x_2^2+9216 x_1^2 x_2^4 x_3^2+9216 x_1^2
x_3^4 x_4^2 \nonumber \\&& +9216 x_1^2 x_4^4 x_3^2+9216 x_1^4
x_3^2 x_4^2+9216 x_2^4 x_3^2 x_4^2+9216 x_2^2 x_3^4 x_4^2
\nonumber \\&& +4608 x_1^4 x_4^4+3072 x_1^6 x_3^2+3072 x_1^6
x_4^2+768 x_1^8+768 x_2^8 \nonumber \\&& +768 x_4^8+768 x_3^8+9216
x_1^4 x_2^2 x_3^2
\end{eqnarray}
Similarly, the corresponding scalar curvature is given by the
following expression
\begin{eqnarray}
 R = -\frac{1}{2}
\frac{1}{(x_1^2+ x_2^2+ x_3^2+ x_4^2)^2}.
\end{eqnarray}
Herewith, we find that the scalar curvature may be described as an
inverse square of the following pure quadratic form $\sum_{i=1}^4
x_i^2$. This shows that the limiting global moduli interactions,
following from the intrinsic geometric scalar curvature, depend
only on the the sum of the squares of the constituent moduli
fields.

For the vanishing value of the Fayet parameter $b=0$, it turns out
that the planar and hyperplanar stabilities are well ensured, in
the limit of equal values of the moduli fields, viz., $x_i=x$. In
particular, we notice, for the case of $b= 0$, that the principle
minors $\{g_2, g_3 \}$ correspond to the following positive
monomial expressions, $g_2:= 512x^4$ and $g_3 = 10240x^6$.
Furthermore, we find, for equal values of the real scalars $x_i=
x$ and vanishing Fayet parameter $b= 0$, that the determinant of
the underlying metric tensor reduces to the following positive
monomial $g= 196608x^8$. Thus, we find that the limiting equal
moduli marginal configuration remains stable under all possible
fluctuations of the moduli $x$.

On the other hand, we see, under the reflection of the Fayet
parameter $b= -a$, that the components of the metric tensor reduce
to the following expressions
\begin{eqnarray}
g_{x_ix_i} &=& 12 x_i^2+4 \sum_{j \neq i} x_j^2-4 a \nonumber \\
g_{x_ix_j} &=& 8 x_i x_j, i \neq j, i,j=1,2,3,4
\end{eqnarray}
Furthermore, it turns out, under the transformation $b=-a$, that
the corresponding planar and hyperplanar stability constraints are
rearranged as per the following general polynomials,
\begin{eqnarray}
 g_2 := 16 a^2-g_{21} a+g_{20}
\end{eqnarray}
\begin{eqnarray}
 g_3 := -64 a^3+g_{32} a^2-g_{31} a+g_{30}
\end{eqnarray}
\begin{eqnarray}
 g := g_{44} a^4- g_{43}a^3+g_{42} a^2-g_{41} a+g_{40}
\end{eqnarray}
In this case, we notice that the scalar curvature transforms into
the following expression
\begin{eqnarray}
R:= \frac{9}{2}\frac{(\sum_{i=1}^4 x_i^2)}{(a- 3\sum_{i=1}^4
x_i^2)} \frac{1}{(a^2-r_{m1} a+r_{m0})}
\end{eqnarray}

Under the inverse of the Fayet parameter $ b:=1/c$, we observe
that the local pair correlations scale as
\begin{eqnarray}
g_{x_ix_i} &=& \frac{4}{c} ((3 x_i^2+ \sum_{i \neq j} x_j^2) c+1) \nonumber \\
g_{x_ix_j} &=& 8 x_i x_j, \ \forall i\neq j, i, j= 1,2,3,4
\end{eqnarray}
Thus, the inverse transformation $b = 1/c$ indicates that the
corresponding values of the principle minors transform according
to the root-set of the underlying polynomial equations. In this
case, we see that the principle minors get inverted in the Fayet
parameter, viz., the associated coefficients of the principle
minors get symmetrically exchanged in those of the concerned
polynomial expressions. In particular, we find that the principle
minors transform according to
\begin{eqnarray}
 g_2 := \frac{1}{c^2} (g_{22}+g_{21}c+g_{20}c^2)
\end{eqnarray}
\begin{eqnarray}
 g_3 :=\frac{1}{c^3}(g_{33}+g_{32}c+g_{31}c^2+g_{30}c^3)
\end{eqnarray}
\begin{eqnarray}
 g= \frac{1}{c^4}(g_{44}+g_{43}c+g_{42}c^2+g_{41}c^3+g_{40}c^4)
\end{eqnarray}
Under the inversion of the Fayet parameter, we may notice further
that there exists a similar scaling property for the associated
intrinsic scalar curvature. We observe that the scalar curvature
transforms into the following expression
\begin{eqnarray}
R = -\frac{9}{2} c^3 \frac{\sum_{i=1}^4x_i^2}{(1+3c
\sum_{i=1}^4x_i^2)}\frac{1}{(r_{m0} c^2+r_{m1}c+1)},
\end{eqnarray}
where the coefficients $r_{m0}$ and $r_{m0}$ remain unchanged,
under the dilatation $b=1/c$.
\section{Threshold Stability}
In this section, we shall analyze threshold stability conditions
and thereby provide an account of the BPS configuration, which can
decay to an intriguing set of other moduli configurations. From
the perspective of the supergravity constraints, it is known that
the charges of the theory must be chosen, such that the effective
potential vanishes at least for some specific value of the Fayet
parameter. Thus, one needs to choose at least two abelian scalar
fields, in order to preserve some supersymmetry. Before proceeding
to the general consideration, we shall examine an associated
interesting threshold configuration of two $U(1)$ scalars, which
involves only the two real scalar fields. In the next subsection,
we shall offer stability properties for such a configuration of
the moduli fields.

\subsection{Simple Complex Scalars}
Let us consider two complex scalar fields $\{ \varphi_1,
\varphi_2\}$ and focus our attention on the specific subsector of
the moduli configuration which involves only the two real scalars.
We may easily see that the general choice of the present interest
may be defined as an element $\{ \varphi_1, \varphi_2\} \in
\mathcal{M}$, such that the $D$-term potential vanishes at the
vacuum. In fact, it is not difficult to see that the possible
values of the fields are an element of $\mathcal{M}$ which may, in
extreme, be defined as the following set,
\begin{eqnarray}
\mathcal M:= \{ \{(x_1,0),(x_2,0)\},  \{(x_1,0),(0, x_2)\}, \{(0,
x_1),(x_2,0)\}, \{(0, x_1),(0, x_2)\} \}
\end{eqnarray}
In order to ensure supersymmetry, we may be required to choose two
alternating abelian charges, which in an appropriate normalization
correspond to $q_1= 1$ and $q_2= -1$. Whence, the threshold
stability of the $D$-term moduli configurations may be specified
by the following Fayet like potential,
\begin{eqnarray}
V(x_1,x_2) := (x_1^2-x_2^2+b)^2
\end{eqnarray}
The components of the covariant metric tensor may thus be
expressed as,
\begin{eqnarray}
g_{x_1x_1}&=& 12 x_1^2-4 x_2^2+4 b \nonumber \\
g_{x_1x_2} &=& -8 x_2 x_1 \nonumber \\
g_{x_2x_2} &=& 12 x_2^2-4 x_1^2-4 b
\end{eqnarray}
It is straightforward to show that the determinant of the metric
tensor is
\begin{eqnarray}
g = -16(b^2- 4(x_1^2- x_2^2) b- 3(x_1^2- x_2^2)^2)
\end{eqnarray}
As in the case of the marginally stable configurations, we find in
the present case that the Ricci scalar takes the following form,
\begin{eqnarray}
R = -b \frac{(x_1^2-x_2^2)}{(b^2- 4(x_1^2- x_2^2) b- 3(x_1^2-
x_2^2)^2)^2}
\end{eqnarray}
Notice that the scalar curvature involves only even powers of the
moduli fields. Thus, the global moduli interactions remain intact
under the reflection of the moduli fields $\{x_1, x_2\}$. For the
case of the vanishing Fayet parameter $b= 0$, we have the
following local pair correlations
\begin{eqnarray}
g_{x_1x_1}&=& 12 x_1^2-4 x_2^2 \nonumber \\
g_{x_1x_2}&=& -8 x_2 x_1 \nonumber \\
g_{x_2x_2}&=& 12 x_2^2-4 x_1^2
\end{eqnarray}
From the above equations, we observe that the components of the
metric tensor are symmetric under the exchange of the moduli
fields, viz., we have $g_{x_1x_1}= g_{x_2x_2}$, under the
replacement of the moduli field $x_1$ by $x_2$, and vice-versa.
Furthermore, the determinant of the metric tensor reduces to a
pure quartic form: $g = 96x_1^2x_2^2 - 48x_1^4 - 48x_2^4$, while
the scalar curvature vanishes identically. In this case, we find a
series of striking notions of the wall crossing pertaining to the
$D$-term potentials. Specifically, the value of the Fayet
parameter $b$ ascribes the nature of the walls of stability as the
polynomial invariance of the intrinsic geometric configuration.

On the other hand, we see, for equal values of the real scalars
$x_1= x_2= x$, that the components of the metric tensor reduce to
the following asymmetric local correlations
\begin{eqnarray}
g_{x_1x_1}&=& 8 x^2+4 b \nonumber \\
g_{x_1x_2}&=& -8 x^2 \nonumber \\
g_{x_2x_2} &=& 8 x^2-4 b
\end{eqnarray}
It turns out that the above asymmetry leads to an unstable
configuration of the moduli fields. The reason follows from the
fact that the underlying determinant of the metric tensor reduces
to a negative value $g = -16b^2$. In this case, we notice further
that the scalar curvature vanishes identically. It may thus be
envisaged, for equal values of the moduli fields $x_1= x_2= x$,
that the $D$-term moduli configurations become uniformly unstable
and statistically non-interacting, during the limiting threshold
transitions in the moduli fields $\{x_1, x_2\}$.

Under the reflection of the Fayet parameter, viz., $b=-a$, we find
that the determinant of the metric tensor reduces to
\begin{eqnarray}
g = -16(a^2-(4 x_1^2-4 x_2^2) a-6 x_2^2 x_1^2+3 x_1^4+3 x_2^4)
\end{eqnarray}
while the scalar curvature reads
\begin{eqnarray}
R = a \frac{(x_1^2-x_2^2)}{(a^2-(4 x_1^2-4 x_2^2) a-6 x_2^2
x_1^2+3 x_1^4+3 x_2^4)^2}
\end{eqnarray}
under the reflection transformation $b = -a$.

\begin{figure}
\hspace*{0.5cm}
\includegraphics[width=8.0cm,angle=-90]{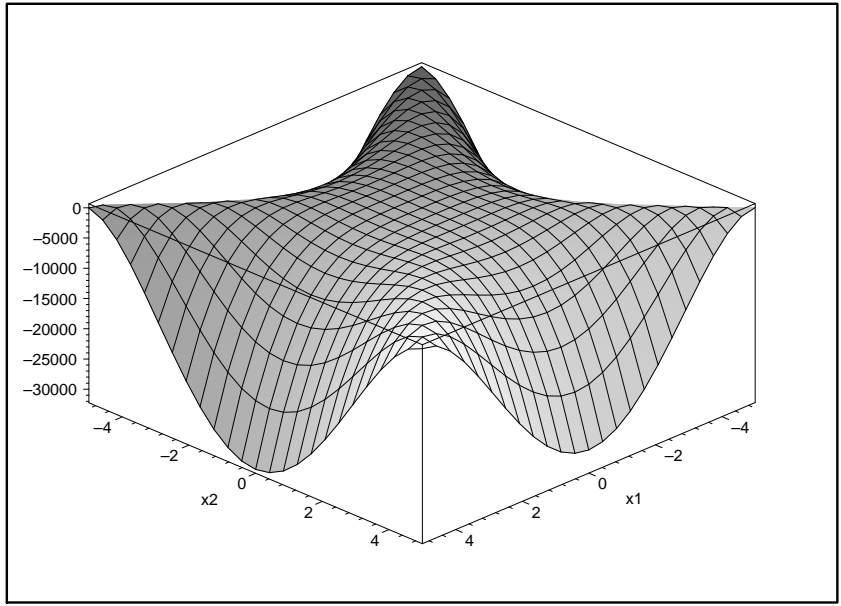}
\caption{The determinant of the metric tensor plotted as a
function of the moduli fields, $x_1, x_2$, describing the
fluctuations in the marginal configuration.} \label{threshold1g}
\vspace*{0.5cm}
\end{figure}

\begin{figure}
\hspace*{0.5cm}
\includegraphics[width=8.0cm,angle=-90]{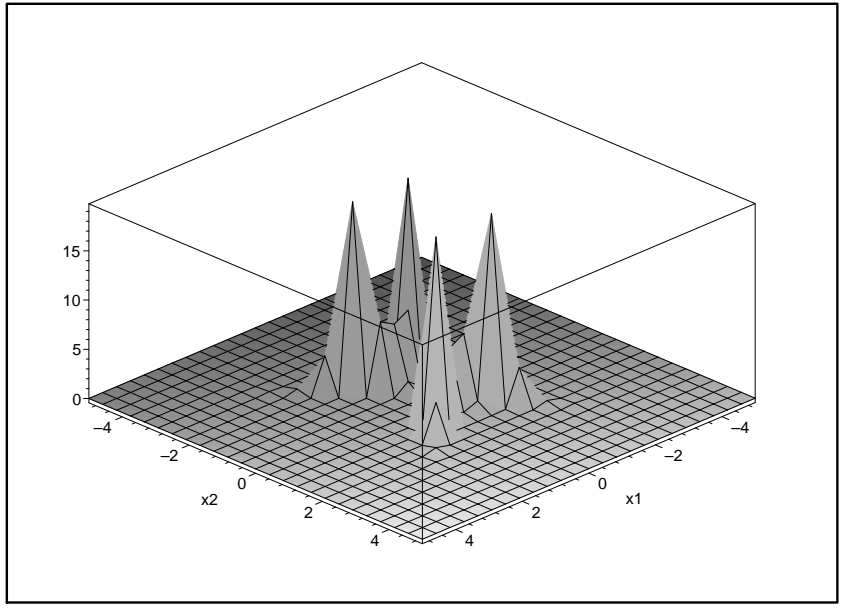}
\caption{The scalar curvature plotted as a function of the moduli
fields, $x_1, x_2$, describing the fluctuations in the marginal
configuration.} \label{threshold1R} \vspace*{0.5cm}
\end{figure}

Under the inversion of the Fayet parameter, viz., $b= 1/c$, it is
easy to observe that the determinant of the metric tensor may
easily be expressed as,
\begin{eqnarray}
g = -\frac{16}{c^2} (3 (x_1^2- x_2^2)^2 c^2+4(x_1^2- x_2^2) c+1)
\end{eqnarray}
The respective value of the scalar curvature is given by
\begin{eqnarray}
R = -c^3\frac{(x_1^2-x_2^2)}{((3 (x_1^2- x_2^2)^2 c^2+4(x_1^2-
x_2^2) c+1))^2}
\end{eqnarray}
We thus notice that the dilatation of $b$ keeps the singularity
structure intact, except for the case when the Fayet parameter
vanishes, viz., we have avoided the inversion of the limit $b=0$.
The graphical viewpoint of the moduli interaction is depicted in
Figs.(\ref{threshold1g}, \ref{threshold1R}). In order to make a
qualitative comparison with the outcomes of the previous section,
the figures of the present interests are also plotted for the unit
Fayet parameter $b=1$.

\subsection{General Complex Scalars}
In this subsection, we shall provide a detailed analysis for the
most general two complex moduli configuration involving two $U(1)$
fields and, in general, four real scalar fields. Firstly, we shall
concentrate on the two complex scalar fields $\{ \varphi_1,
\varphi_2 \in U(1) \}$ and supply an intriguing stability
analysis, in terms of respective real scalars $(x_1, x_2, x_3,
x_4)$. Furthermore, it turns out that one may as well analyze the
nature of D-terms threshold stability configurations for the set
of general moduli $\{ \varphi_1, \varphi_2, ..., \varphi_n \}$.
Nevertheless, the threshold stability of general two charge BPS
configurations may be investigated by prescribing four real scalar
fields, $x_a= (x_1, x_2,x_3, x_4)$ ascribing the D-term potential,
\begin{eqnarray}
 V(x_1,x_2,x_3,x_4) := (x_1^2+x_2^2-x_3^2-x_4^2+b)^2
\end{eqnarray}

For the given moduli fields $(x_1, x_2,x_3, x_4)$, it is however
not difficult to show that the components of the covariant metric
tensor may be expressed as,
\begin{eqnarray}
g_{x_ix_i} &=& 12 x_i^2+4 x_j^2-4\sum_{k\neq i,j} x_k^2+4 b, i,j=1,2  \nonumber \\
g_{x_ix_i} &=& 12 x_i^2+4 x_j^2-4\sum_{k\neq i,j} x_k^2-4 b, i,j=3,4  \nonumber \\
g_{x_ix_j} &=& 8 x_i x_j, (i,j) \in \{(1,2),(3,4)\} \nonumber \\
g_{x_ix_j} &=& -8 x_i x_j, (i,j) \in \{(1,3),(1,4),(2,3)(2,4)\} 
\end{eqnarray}

The planar and hyper planar stability may now be analyzed, as
well. In order to do so, we would need to consider the principle
minors of the underlying intrinsic metric tensor. In fact, one may
find that the planar principle minor, defined as $g_2:=
g_{11}g_{22}- g_{22}^2$, leads to the following quadratic
polynomial
\begin{eqnarray}
g_2:=  g_{22}b^2+ g_{21}b+ g_{20}
\end{eqnarray}
where the coefficients are given by the combinations
\begin{eqnarray}
g_{22}&:=&16  \nonumber \\
g_{21}&:=&(64 x_2^2-32 x_3^2+64 x_1^2-32 x_4^2) \nonumber \\
g_{20}&:=&64 x_1^2 x_3^2+96 x_2^2 x_1^2+16 x_4^4-64 x_2^2 x_4^2+48
x_2^4 \nonumber \\ && +16 x_3^4+32 x_3^2 x_4^2+48 x_1^4-64 x_2^2
x_3^2-64 x_1^2 x_4^2
\end{eqnarray}

Furthermore, a straightforward computation thence shows that the
hyper-planar principle minor, defined as $g_3 =
g_{11}(g_{22}g_{33}- g_{22}^2) g_{12}(g_{12}g_{33}- g_{13}g_{23})+
g_{13}(g_{12}g_{23}- g_{13}g_{22})$, reduces to the following
cubic polynomial
\begin{eqnarray}
g_3 := g_{33}b^3+ g_{32} b^2+ g_{31}b +g_{30}
\end{eqnarray}
where the coefficients of the cubic equation are given by the
expressions
\begin{eqnarray}
g_{33}&:=& -64 \nonumber \\
g_{32}&:=&320 x_3^2+192 x_4^2-320 x_1^2-320 x_2^2 \nonumber \\
g_{31}&:=&896 x_2^2 x_3^2-448 x_3^4+640 x_2^2 x_4^2-448 x_2^4
\nonumber \\ && -448 x_1^4+896 x_1^2 x_3^2+640 x_1^2 x_4^2-192
x_4^4 \nonumber \\ &&
-640 x_3^2 x_4^2-896 x_2^2 x_1^2 \nonumber \\
g_{30}&:=& 896 x_1^2 x_2^2 x_4^2-576 x_1^4 x_2^2+448 x_1^4
x_4^2-576 x_1^2 x_2^4 \nonumber \\ && +1152 x_1^2 x_2^2 x_3^2-896
x_1^2 x_3^2 x_4^2-192 x_2^6+64 x_4^6 \nonumber \\ && +192
x_3^6-192 x_1^6-320 x_1^2 x_4^4-320 x_2^2 x_4^4+576 x_1^4 x_3^2
\nonumber \\ && +448 x_2^4 x_4^2+320 x_3^2 x_4^4-576 x_1^2
x_3^4+576 x_2^4 x_3^2 \nonumber \\ && -576 x_2^2 x_3^4-896 x_2^2
x_3^2 x_4^2+448 x_3^4 x_4^2
\end{eqnarray}

It is not difficult to observe that the determinant of the metric
tensor is given by the following quartic polynomial
\begin{eqnarray}
 g := g_{44} b^4+ g_{43}b^3+g_{42} b^2+g_{41} b+g_{40}
\end{eqnarray}
where the coefficients appearing in the above equation read
\begin{eqnarray}
g_{44}&:=& 256  \nonumber \\
g_{43}&:=& 1536 x_2^2-1536 x_3^2-1536 x_4^2+1536 x_1^2  \nonumber \\
g_{42}&:=& 3072 x_4^4+3072 x_1^4-6144 x_1^2 x_4^2+3072 x_2^4
\nonumber \\ && +3072 x_3^4-6144 x_2^2 x_4^2-6144 x_1^2 x_3^2+6144
x_3^2 x_4^2 \nonumber \\ &&
-6144 x_2^2 x_3^2+6144 x_2^2 x_1^2  \nonumber \\
g_{41}&:=& -15360 x_1^2 x_2^2 x_3^2-2560 x_3^6+7680 x_2^2
x_3^4-15360 x_1^2 x_2^2 x_4^2  \nonumber \\ && -2560 x_4^6+2560
x_2^6-7680 x_2^4 x_4^2-7680 x_2^4 x_3^2-7680 x_1^4 x_4^2 \nonumber
\\ && +7680 x_1^2 x_2^4-7680 x_3^4 x_4^2+7680 x_1^2 x_4^4+7680
x_1^2 x_3^4  \nonumber \\ && +7680 x_1^4 x_2^2+7680 x_2^2
x_4^4+2560 x_1^6-7680 x_3^2 x_4^4  \nonumber \\ &&
+15360 x_2^2 x_3^2 x_4^2+15360 x_1^2 x_3^2 x_4^2-7680 x_1^4 x_3^2  \nonumber \\
g_{40}&:=& 4608 x_3^4 x_4^4+9216 x_2^4 x_4^2 x_3^2-9216 x_2^2
x_3^4 x_4^2  \nonumber \\ && -9216 x_2^2 x_4^4 x_3^2+768
x_3^8+3072 x_3^6 x_4^2-3072 x_3^6 x_1^2 \nonumber \\ && -3072
x_3^6 x_2^2-3072 x_1^6 x_4^2+3072 x_1^6 x_2^2-3072 x_1^6 x_3^2
\nonumber \\ && -3072 x_2^6 x_4^2+3072 x_2^6 x_1^2-3072 x_2^6
x_3^2-3072 x_4^6 x_1^2  \nonumber \\ && -3072 x_4^6 x_2^2+3072
x_4^6 x_3^2+4608 x_1^4 x_3^4+4608 x_1^4 x_2^4 \nonumber \\ &&
+4608 x_1^4 x_4^4+4608 x_2^4 x_4^4+4608 x_2^4 x_3^4-9216 x_1^2
x_2^4 x_3^2 \nonumber \\ && +9216 x_1^2 x_2^2 x_3^4-9216 x_1^2
x_3^2 x_4^4+9216 x_1^4 x_3^2 x_4^2  \nonumber \\ && -9216 x_1^2
x_3^4 x_4^2-9216 x_1^4 x_2^2 x_3^2+9216 x_1^2 x_2^2 x_4^4
\nonumber \\ && -9216 x_1^4 x_2^2 x_4^2-9216 x_1^2 x_2^4 x_4^2+768
x_1^8+768 x_2^8 \nonumber \\ && +768 x_4^8+18432 x_2^2 x_1^2 x_4^2
x_3^2
\end{eqnarray}
It is not difficult to show that the Ricci scalar curvature takes
the intriguing form
\begin{eqnarray}
R = -\frac{9}{2} \frac{x_1^2-x_3^2-x_4^2+x_2^2}{(b+3 x_1^2-3
x_3^2+3 x_2^2-3 x_4^2)} \frac{1}{(b^2+ r_{m1} b+ r_{m0})}
\end{eqnarray}
where the moduli functions $r_{m1}$ and $r_{m0}$ can be defined as
\begin{eqnarray}
r_{m1}&:=& (4 x_1^2+4 x_2^2-4 x_3^2-4 x_4^2)  \nonumber \\
r_{m0}&:=& (+3 x_1^4+3 x_4^4-6 x_1^2 x_3^2-6 x_1^2 x_4^2+6 x_1^2
x_2^2  \nonumber \\&& +6 x_3^2 x_4^2-6 x_2^2 x_3^2-6 x_2^2 x_4^2+3
x_2^4+3 x_3^4)
\end{eqnarray}

\begin{figure}
\hspace*{0.5cm}
\includegraphics[width=8.0cm,angle=-90]{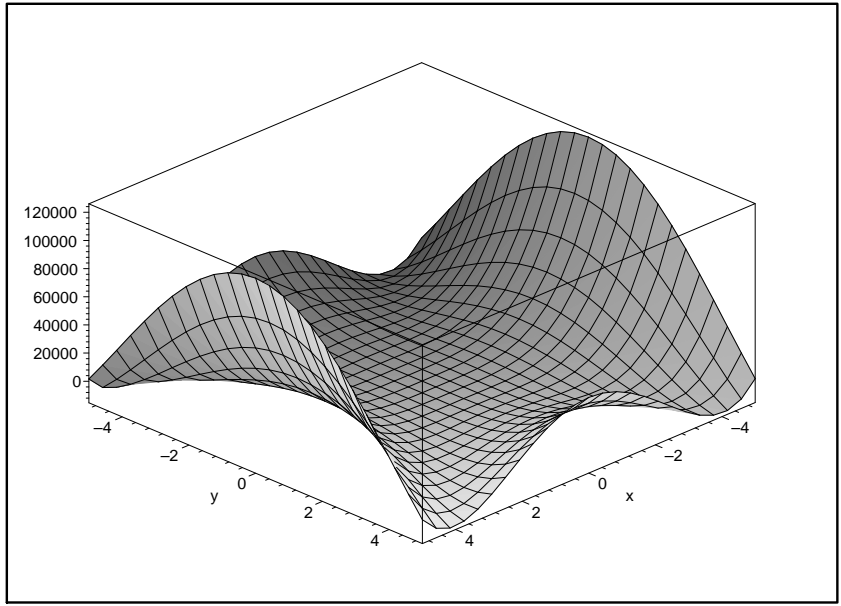}
\caption{The surface minor plotted as a function of the moduli
fields, $x, y$, describing the fluctuations in the threshold
configuration.} \label{threshold2g2} \vspace*{0.5cm}
\end{figure}

\begin{figure}
\hspace*{0.5cm}
\includegraphics[width=8.0cm,angle=-90]{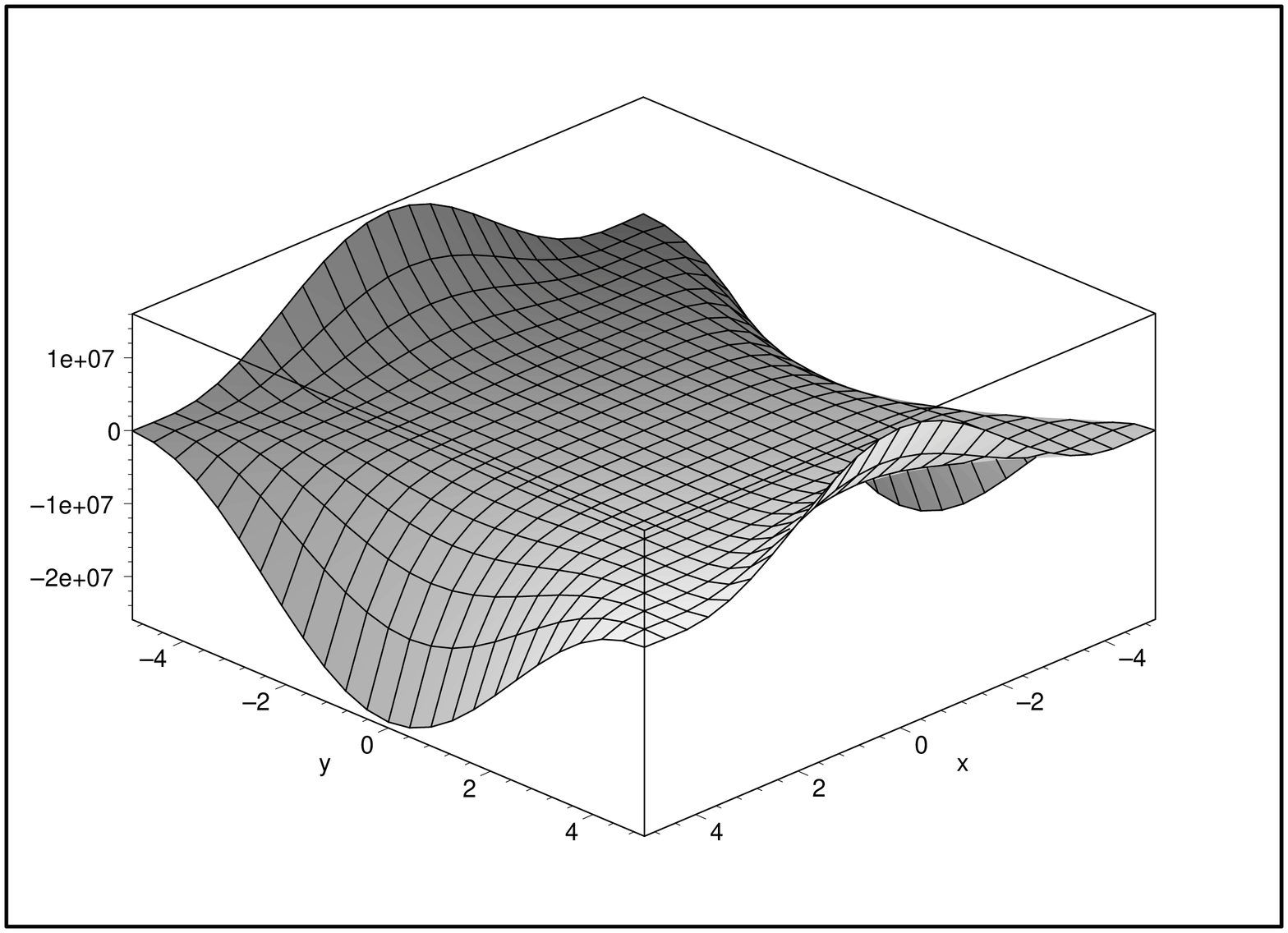}
\caption{The hypersurface minor plotted as a function of the
moduli fields, $x, y$, describing the fluctuations in the
threshold configuration.} \label{threshold2g3} \vspace*{0.5cm}
\end{figure}

\begin{figure}
\hspace*{0.5cm}
\includegraphics[width=8.0cm,angle=-90]{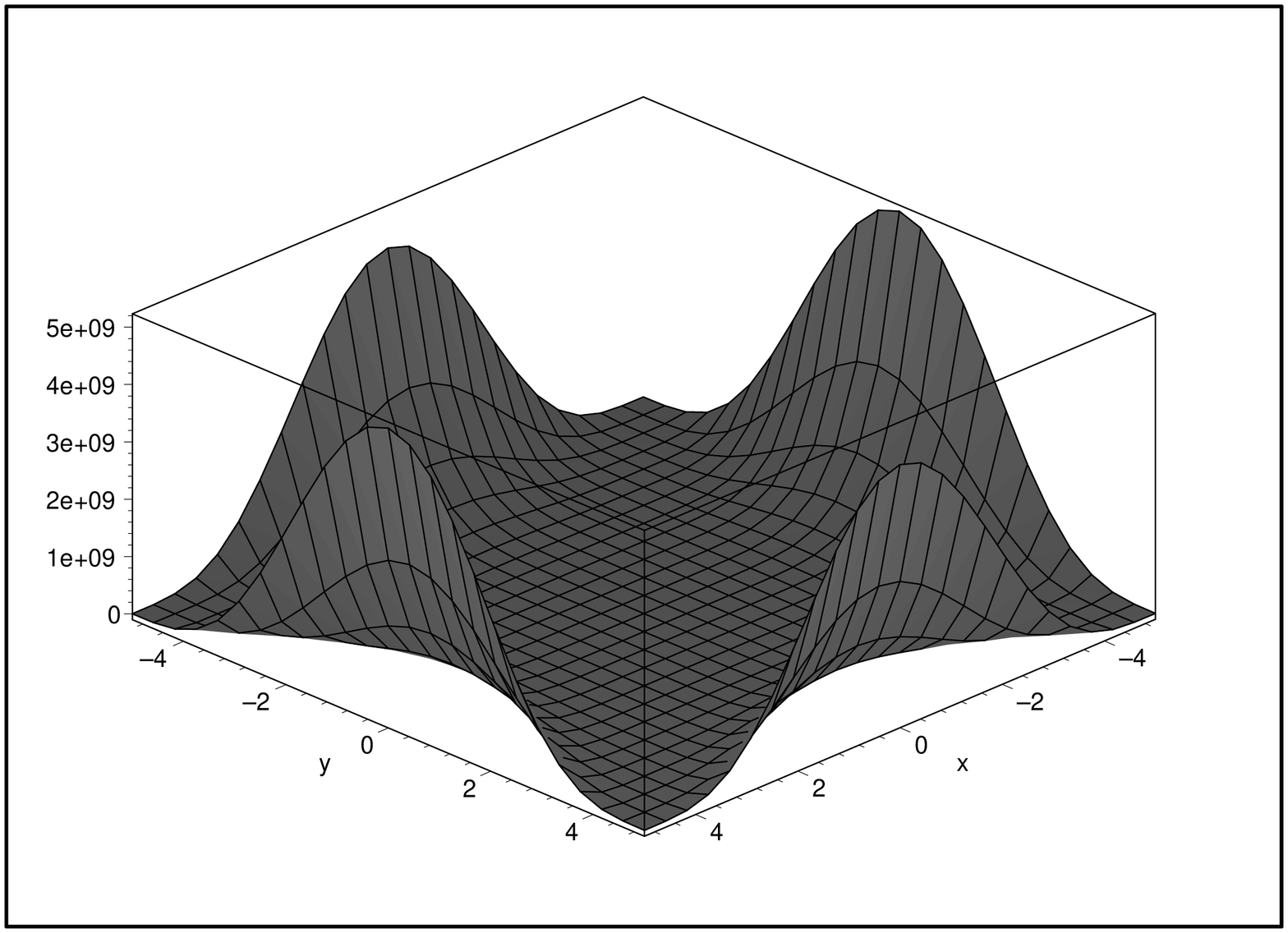}
\caption{The determinant of the metric tensor plotted as a
function of the moduli fields, $x, y$, describing the fluctuations
in the threshold configuration.} \label{threshold2g}
\vspace*{0.5cm}
\end{figure}

\begin{figure}
\hspace*{0.5cm}
\includegraphics[width=8.0cm,angle=-90]{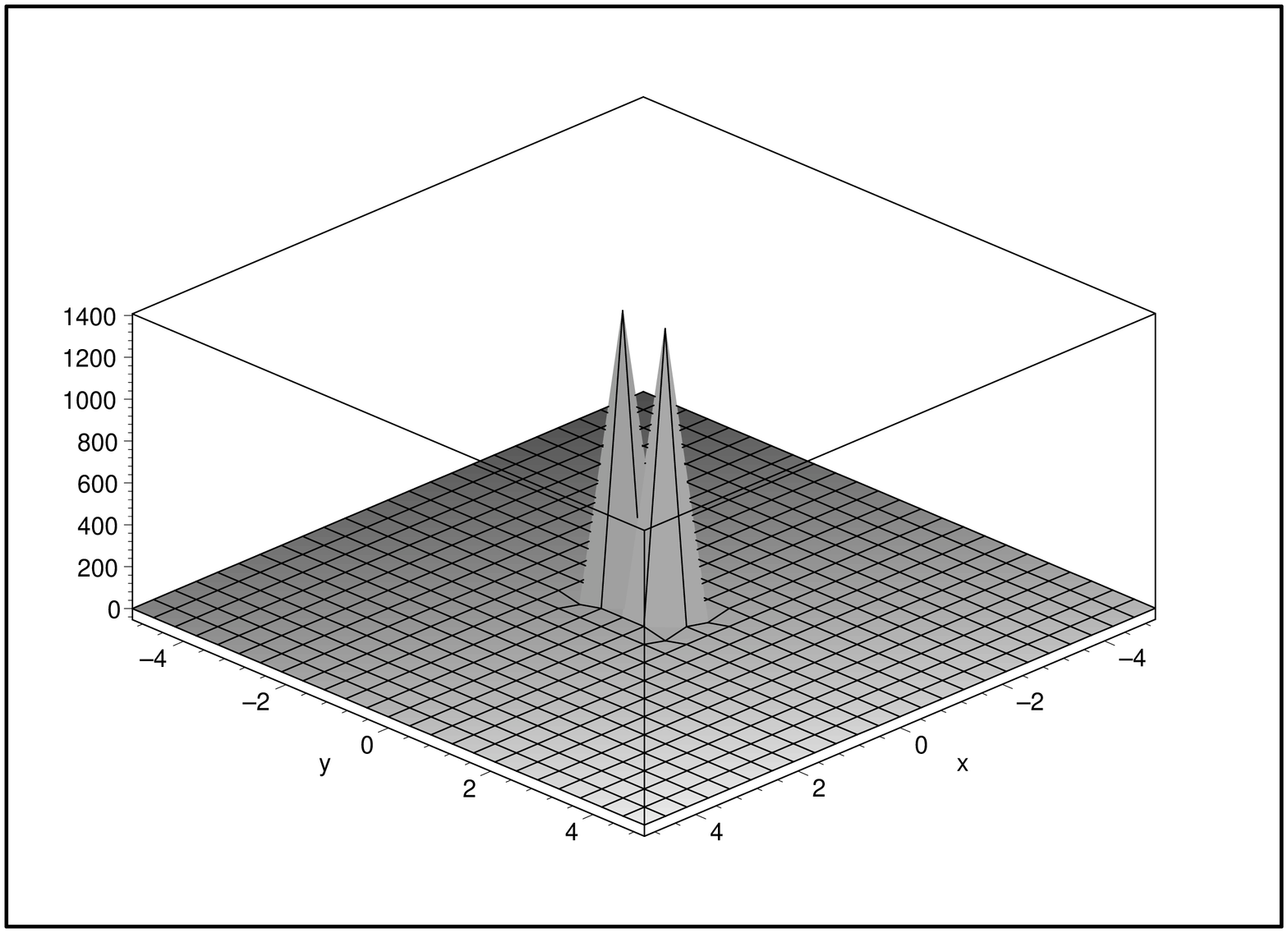}
\caption{The scalar curvature plotted as a function of the moduli
fields, $x, y$, describing the fluctuations in the threshold
configuration.} \label{threshold2R} \vspace*{0.5cm}
\end{figure}

For a pairwise equality: $x_1= x_2=  x$, $x_3= x_4=  y$ and with
the unit Fayet parameter $b=1$, the principle minors are given by
\begin{eqnarray}
g_2 &=& -256 x^2 y^2+192 x^4+16+128 x^2-64 y^2+64 y^4
\end{eqnarray}
\begin{eqnarray}
g_3 &=& -64-1792 x^4-1280 y^4+3072 x^2 y^2-640 x^2+512 y^2
\nonumber \\ && +4096 x^4 y^2-3584 y^4 x^2+1024 y^6-1536 x^6
\end{eqnarray}
\begin{eqnarray}
g &=& 256+12288 x^4+12288 y^4-24576 x^2 y^2+3072 x^2 \nonumber \\
&& -3072 y^2-61440 x^4 y^2+73728 x^4 y^4-49152 x^6 y^2 \nonumber
\\ && -49152 y^6 x^2+61440 y^4 x^2+12288 x^8+12288 y^8 \nonumber
\\ && -20480 y^6+20480 x^6
\end{eqnarray}
In this case, the associated scalar curvature is given by
\begin{eqnarray}
 R = -9 \frac{(x^2- y^2)}{(1+6 x^2-6 y^2)(1+12 x^4+8 x^2-24 x^2 y^2+12 y^4-8 y^2)}
 \end{eqnarray}
The behavior of the principle minors $g_2, g_3$, the determinant
of the metric tensor $g$ and the corresponding scalar curvature
$R$ have been respectively shown in the Figs.(\ref{threshold2g2},
\ref{threshold2g3}, \ref{threshold2g}, \ref{threshold2R}).

For equal values of the scalar fields, viz, $x=z$, $y=z$ 
%
the respective limiting values of the principle minors and
determinant of the metric tensor are given by
\begin{eqnarray}
 g_2 := 16 b (b+4 x^2)
\end{eqnarray}
\begin{eqnarray}
 g_3 := -64 b^2 (b+2 x^2)
\end{eqnarray}
\begin{eqnarray}
 g := 256 b^4
\end{eqnarray}
In this case, we find that the associated scalar curvature
vanishes identically, viz., for all values of the Fayet parameter,
we have a non-interacting statistical configuration with the
following scalar curvature
\begin{eqnarray}
 R = 0
\end{eqnarray}

Interestingly, we note, for the limiting threshold configuration
with the vanishing Fayet parameter $b = 0$, that the stabilities
on a line, surface, hyper-surface, as well as the stability of the
entire moduli configuration, are respectively described by the
properties of polynomials of degree 2, 4, 6 and 8. Specifically,
for the limiting Fayet parameter $b=0$, we find, from the
perspective of the intrinsic geometry, that the diagonal and
off-diagonal pair correlations reduce to a set of quadratic
polynomials and monomials. For example, the first diagonal and the
first off-diagonal components reduce to the following expressions
\begin{eqnarray}
g_{x_1x_1} &=& 12 x_1^2+4 x_2^2-4 x_3^2-4 x_4^2 \nonumber \\
g_{x_1x_2} &=& 8 x_1 x_2
\end{eqnarray}
The surface minor reduces to the following quartic polynomial
\begin{eqnarray}
g_2 &=& -64 x_1^2 x_3^2+96 x_2^2 x_1^2+16 x_4^4-64 x_2^2 x_4^2+48
x_2^4  \nonumber \\&& +16 x_3^4+32 x_3^2 x_4^2+48 x_1^4-64 x_2^2
x_3^2-64 x_1^2 x_4^2
\end{eqnarray}
The question of the stability of the hypersurface of the moduli
configuration may be determined by the associated minor of the
metric tensor. It follows that the hypersurface minor is given by
\begin{eqnarray}
g_3 &=& 896 x_1^2 x_2^2 x_4^2-576 x_1^4 x_2^2+448 x_1^4 x_4^2-576
x_1^2 x_2^4  \nonumber \\&& +1152 x_1^2 x_2^2 x_3^2-896 x_1^2
x_3^2 x_4^2-192 x_2^6+64 x_4^6  \nonumber \\&& +192 x_3^6-192
x_1^6-320 x_1^2 x_4^4-320 x_2^2 x_4^4+576 x_1^4 x_3^2  \nonumber
\\&& +448 x_2^4 x_4^2+320 x_3^2 x_4^4-576 x_1^2 x_3^4+576 x_2^4
x_3^2  \nonumber \\&& -576 x_2^2 x_3^4-896 x_2^2 x_3^2 x_4^2+448
x_3^4 x_4^2
\end{eqnarray}
Finally, an existence of the global stability of the vanishing
Fayet parameter threshold moduli configuration may be determined
by the positivity of the determinant of the underlying metric
tensor. In this limit, we find that the determinant of the metric
tensor takes the following homogeneous polynomial form
\begin{eqnarray}
g &=& 9216 x_1^4 x_3^2 x_4^2+9216 x_2^4 x_4^2 x_3^2+9216 x_1^2
x_2^2 x_4^4  \nonumber \\&& +9216 x_1^2 x_2^2 x_3^4+18432 x_2^2
x_1^2 x_4^2 x_3^2+4608 x_3^4 x_4^4  \nonumber \\&& +3072 x_3^6
x_4^2-3072 x_3^6 x_1^2-3072 x_3^6 x_2^2-3072 x_1^6 x_4^2 \nonumber
\\&& -3072 x_1^6 x_3^2+3072 x_2^6 x_1^2-3072 x_2^6 x_3^2-3072
x_4^6 x_2^2  \nonumber \\&& +3072 x_4^6 x_3^2+768 x_3^8+768
x_1^8+768 x_2^8+768 x_4^8  \nonumber \\&& +4608 x_1^4 x_3^4+4608
x_1^4 x_4^4+4608 x_2^4 x_4^4+4608 x_2^4 x_3^4 \nonumber \\&& +4608
x_1^4 x_2^4-9216 x_1^2 x_2^4 x_3^2-9216 x_1^4 x_2^2 x_3^2
\nonumber \\&& -9216 x_2^2 x_4^4 x_3^2-9216 x_1^2 x_2^4 x_4^2-9216
x_1^2 x_3^2 x_4^4  \nonumber \\&& -9216 x_2^2 x_3^4 x_4^2-9216
x_1^4 x_2^2 x_4^2-9216 x_1^2 x_3^4 x_4^2 \nonumber \\&& +3072
x_1^6 x_2^2-3072 x_2^6 x_4^2-3072 x_4^6 x_1^2
\end{eqnarray}
Moreover, in the limit of $b=0$, we observe that the Ricci scalar
curvature reduces to the inverse square of a unique quadratic
polynomial of the moduli scalars, which defines the $D$-term
potential of the associated threshold configuration. Specifically,
we find that the limiting scalar curvature may be expressed as
\begin{eqnarray}
 R = -\frac{1}{2}
\frac{1}{(x_1^2+x_2^2- x_3^2- x_4^2)^2}
\end{eqnarray}

For the case of the reflected Fayet parameter $b=-a$, the
components of the matric tensor transform as per our general
expectation. For example, the first diagonal and the first
off-diagonal components of the metric tensor transform as follows
\begin{eqnarray}
 g_{x_1x_1} &=& 12 x_1^2+4 x_2^2-4 x_3^2-4 x_4^2-4 a \nonumber \\
 g_{x_1x_2} &=& 8 x_1 x_2,
\end{eqnarray}
Under the reflection of the Fayet parameter $b= -a$, we find that
the following value is satisfied by the planar minor,
\begin{eqnarray}
 g_2 := 16 a^2-g_{21} a+g_{20}
\end{eqnarray}
Nevertheless, it is not difficult to see that the hyper-planar
minor reduces to the following expression
\begin{eqnarray}
 g_3 := -64 a^3+g_{32} a^2-g_{31} a+g_{30}
\end{eqnarray}
Under the reflection of the Fayet parameter, the following
expression gives an intriguing comparison of the determinant of
the metric tensor
\begin{eqnarray}
g := g_{44} a^4- g_{43}a^3+g_{42} a^2-g_{41} a+g_{40}
\end{eqnarray}
Similarly, one may easily check that the scalar curvature
satisfies,
\begin{eqnarray}
R:= \frac{9}{2}\frac{(x_1^2-x_3^2-x_4^2+x_2^2)}{(a- (3 x_1^2+3
x_3^2-3 x_2^2+3 x_4^2)} \frac{1}{(a^2-r_{m1} a+r_{m0})}
\end{eqnarray}

Under the inverse of the Fayet parameter $ b:=1/c$, we observe
that only the diagonal pair correlations dilate, while the
off-diagonal pair correlations remain intact. For instance, the
first diagonal and the first off-diagonal components of the metric
tensor scale as follows
\begin{eqnarray}
g_{x_1x_1} &=& \frac{4}{c} ((3 x_1^2+x_2^2-x_3^2-x_4^2) c+1) \nonumber \\
g_{x_1x_2} &=& 8 x_1 x_2
\end{eqnarray}
In this case, we see that the series of the polynomials, which
define the associated principle minors, get inverted in the Fayet
parameter. Under the inversion $b= 1/c$, we may further observe
that the stability of the configuration can stem from the
following transformation rules. In particular, we find that the
respective stability constraints transform as
\begin{eqnarray}
g_2= \frac{1}{c^2} (g_{22}+g_{21}c+g_{20}c^2)
\end{eqnarray}
\begin{eqnarray}
g_3= \frac{1}{c^3}(g_{33}+g_{32}c+g_{31}c^2+g_{30}c^3)
\end{eqnarray}
\begin{eqnarray}
g= \frac{1}{c^4}(g_{44}+g_{43}c+g_{42}c^2+g_{41}c^3+g_{40}c^4)
\end{eqnarray}
Thus, we find that the stability constraints remain alternating
and scale by the inverse of the Fayet parameter, under its
inversion. A straightforward calculation shows that a similar
scaling property arises for the intrinsic scalar curvature. In
this case, we see that the scalar curvature transforms into the
following expression
\begin{eqnarray}
R = -\frac{9}{2} c^3
\frac{x_1^2-x_3^2-x_4^2+x_2^2}{(1+3c(x_1^2-x_3^2+x_2^2-x_4^2)}\frac{1}{(r_{m0}
c^2+r_{m1}c+1)},
\end{eqnarray}
where the coefficients $r_{m0}$ and $r_{m1}$ remain intact as per
their earlier definition. For equal values of the moduli scalars,
the planar, hyper-planar and the determinant stability constraints
imply that the principle minors and the determinant of the metric
tensor respectively transform as $g_2 = (8x^2 + 4/c)^2- 64x^4$,
$g_3 = -64(1 + 2x^2c)/c^3$ and $g = 256/c^4$. For a given nonzero
$c$, this shows that the equality of the constituent moduli fields
implies a relatively unstable intrinsic hypersurface, pertaining
to the threshold configurations.

We thus see that the determinant of the metric tensor is positive
definite. This suggests that the entire threshold moduli
configuration may be stabilized, when all the constituent fields
are allowed to fluctuate. Furthermore, one may easily notice that
the present investigation provides an existence of the intriguing
genesis of moduli stabilization with a vanishing intrinsic scalar
curvature, which specifically occurs for equal vacuum expectation
values of the constituent moduli scalars. Similar intrinsic
geometric studies may further be investigated, in order to acquire
the stability properties of the moduli configurations.
%
\section{Conclusion and Remarks}

The present paper explores the intrinsic geometric stability
properties for the moduli configurations. We exhibit an
interesting set of implications for the $D$-term potentials. In
this study, we find that there exists a set of real intrinsic
Riemannian geometric stability criteria and thereby provide the
underlying implications for the Fayet models and walls of the
marginal and threshold stability properties pertaining to the
$D$-term moduli configurations. More specifically, we find, for a
set of given complex abelian scalar fields satisfying the
D-constraint, that the various possible stability criteria, viz.,
linear, planar, hyper-planar and the entire moduli space
stabilities, may all be easily accomplished over a domain of the
Fayet parameter. The present paper analyzes the moduli
stabilization problem for the marginal and threshold stable moduli
configurations and thereby explicates that these configurations
are rather stable, under the possible transformations of the Fayet
parameter.

We supply an intrinsic geometric method to investigate the nature
of the marginal and threshold walls of the stabilities.
Furthermore, we have demonstrated that the threshold stable
configurations concentrate on the alternating definition of the
$D$-term potential. Thus, the examination of the underlying
intrinsic configuration depends on the true symmetry of the
underlying moduli potential. Nevertheless, the non-vanishing of
the intrinsic scalar curvature defines the possible range of the
global correlations between the constituent real scalar fields.
Whereas, the vanishing curvature indicates that the underlying
moduli space configuration comes out to be a non-interacting
statistical system. After the vanishing value of the scalar
curvature, the configuration changes the nature of the underlying
global interaction among the scalar fields.

More precisely, the intrinsic scalar curvature defines an expected
volume of the global correlation among the constituent scalar
fields, beyond which the underlying moduli configuration becomes a
non-interacting system and thereby changes its intrinsic geometric
nature. The singularity structure of the intrinsic curvature
determines possible phase transitions, if any, in the vacuum
moduli manifold, and thereby infers about its feasible decay to
the different daughter configurations. It is worth to mention that
the threshold stable configurations do not always entail the
positivity of the principle components of the metric tensor,
higher principle minors and the determinant of the intrinsic
metric tensor. Thus, there does not always exist a well-defined
volume form on the underlying vacuum manifold of the threshold
moduli configurations. This supports an intriguing fact that the
marginally stable configurations swiftly decay to another set of
non-supersymmetric BPS-configurations.

We further find that the marginally stable BPS configurations have
the same sign of the principle components of the metric tensor. In
turn, it follows that the two and four dimensional real intrinsic
configurations have a uniform sign for the all components and are
positive definite. Thus, the above moduli configurations are
stable for all non-negative values of the Fayet parameter. In
general, we notice that very similar conclusions are noticed for
both the leading order potentials describing decaying BPS
configurations. In particular, we have shown that the marginal
configurations, with the unit $U(1)$ charge $q_i = 1, \forall i $,
possess the same intrinsic geometric metric tensor. The nature of
principle minors and scalar curvature have also an expected
promising feature. The general properties of the above geometric
quantities remain alternating, for the threshold configurations
defined with the charges $q_i = 1, q_{i+1}= -1 \forall i$. As
mentioned before, such moduli configurations may be piecewise
analyzed. From the perspective of the intrinsic geometry, we
envisage that the present analysis may be extended for a variety
of moduli configurations, e.g., $D$-term and $F$-terms potentials.

Importantly, the subleading corrections nevertheless need be
carefully explored, in order to have a more precise evaluation of
the symmetries of the general moduli configurations, which are
entangled with the invariants of the intrinsic geometry. In
particular, it may be important to distinguish whether the
underlying reflection symmetry and possible scalings are
preserved, under the higher derivative contributions. This task is
left for a future investigation. As mentioned earlier, it is worth
mentioning that a moduli configuration is said to be stable, if
the basic intrinsic geometric elements, \textit{viz.}, the
diagonal components of the metric tensor $g_{ii}$, principle
minors $g_{i}$ and the determinant of the metric tensor $g$ are
positive definite $\forall i \in \Lambda$, constituting the
stability in each possible direction of the chosen moduli
configuration. For a finite range of the Fayet parameter, the
framework of the present approach supports that the general moduli
configurations may be stabilized in a finite domain of the moduli
fields.

We may further observe that the principle components of the metric
tensor retain the same polynomial form in the constituent scalar
fields, while the off-diagonal components turn out to be some
quadratic monomials. The present investigation shows that the
definite character of the intrinsic geometry remains true, for an
arbitrary number of scalar fields satisfying the leading order
$D$-term potential. For the threshold moduli configurations, the
intrinsic scalar curvature, thus investigated, vanishes
identically for the vanishing value of the Fayet parameter.
Whilst, it is not difficult to show that similar facts do not
continue to hold for the underlying determinant of metric tensors.
In particular, we discover that the null value of the Fayet
parameter does not cause the determinants to vanish. Consequently,
our computation demonstrates that the marginally stable
configurations are intrinsic geometrically curved, even for the
vanishing value of the Fayet parameter, whilst the same conclusion
does not hold for threshold stable moduli configurations. In other
words, it may be generically anticipated, for the threshold moduli
configurations, that the intrinsic scalar curvature vanishes
identically for the limiting Fayet parameter $b = 0$. Thus, the
threshold configurations become a non-interacting statistical
system, without the contribution of the Fayet term.

Furthermore, faithfully equivalent conclusions remain true in
common, for equal vacuum values of the scalar fields. In
particular, we find that the marginally stable moduli
configurations are interacting with an underlying intrinsic
curvature $R(x_i)\neq 0$, for identical values of the constituent
scalars $x_i = x$. Whilst, we find for the vanishing Fayet
parameter that the corresponding scalar curvature $R(x_i)$
vanishes identically for the threshold stable configurations,
pertaining to the $D$-term potential and the associated Fayet
model. An inference may thus be made that the marginal
configurations do not change the nature of the intrinsic geometry,
under an increment of vacuum scalar fields, whereas the threshold
configurations do change their intrinsic geometric
characteristics. The present analysis thus offers a stimulating
feature for the possible decaying supergravity configurations.
Specifically, the precise notions, as obtained in the present
case, may shed light on the symmetry properties of the intrinsic
geometric invariant arising from the generic D-term and other
moduli space potentials.

Finally, the divergence structure of the intrinsic curvature
implies possible phase transitions, in the chosen physical vacuum
moduli configurations. As per the guidelines of the present
consideration, such a characterization may be furnished by a
finite set of abelian scalar fields. Although the present analysis
is of leading order in character, however it provides a general
real intrinsic geometric covariant technology, which could be
applied towards stability examinations of the various moduli
configurations, pertaing to the gauge and string theories. In
particular, it is worth mentioning that the moduli space
stabilization problem may be appreciated, from the outset of the
intrinsic real Riemannian geometry. In this perspective, the
potential application may be explored for the determination of
local and global moduli stability domains, involving a class of
moduli configurations of the present experimental and
phenomenological interests of the subject. As mentioned in the
introduction, some of these include Higgs moduli, Calabi-Yau
moduli, torus moduli, instanton moduli, topological string moduli,
and the moduli pertaining to $D$ and $M$-particles. In order to
encompass an explicit stability feature emerging from the present
setup, the concept of the intrinsic geometry may herewith be
brought out into a close connection with the underlying
supergravity configurations. From the perspective of the present
investigation, it may therefore be anticipated that a set of
similar observations would emerge further, for a class of generic
moduli space configurations. This problem is left for a future
consideration.

\section*{Acknowledgments}
This work has been supported in part by the European Research
Council grant n.~226455, \textit{``SUPERSYMMETRY, QUANTUM GRAVITY
AND GAUGE FIELDS (SUPERFIELDS)"}. This work was undertaken during
the \textit{``Strings-2009"}, organized by the \textit{``String
Theory Group of the Physics Department and the INFN Section of the
University of Rome Tor Vergata"}. The nice hospitality of the
"Aula Magna" of the \textit{``Angelicum: The Pontificia
Universit\`{a} S. Tommaso, Roma, Italy"} is cordial appreciated.
The research of BNT was supported in part by the \textit{``CSIR,
New Delhi, India''}; \textit{``Indian Institute of Technology
Kanpur, Uttar Pradesh, India''}, the \textit{``Abdus Salam ICTP"}
contribution to the young researchers (PhD students and young
post-docs from developing countries) to participate in the
\textit{``Strings 2009''} and \textit{``INFN-Laboratori Nazionali
di Frascati, Roma, Italy''}.


\begin{thebibliography}{99}

\bibitem{1a} P. S. Aspinwall, B. R. Greene, D. R. Morrison,
``Measuring Small Distances in N = 2 Sigma Models", Nucl. Phys. B
{\bf 420} (1994) 184-242, {\tt arXiv:hep-th/9311042v1}.

\bibitem{1b} E. Witten, ``String Theory Dynamics In Various Dimensions", Nucl. Phys.
B {\bf 443} (1995) 85-126, {\tt arXiv:hep-th/9503124v2}.

\bibitem{2a} B. R. Greene, J. Levin, ``Dark Energy and Stabilization of Extra
Dimensions", JHEP {\bf 0711}, 096, 2007, {\tt arXiv:0707.1062v2
[hep-th]}.

\bibitem{2b} D. Lust, ``Seeing through the String Landscape - a
String Hunter's Companion in Particle Physics and Cosmology", JHEP
{\bf 0903} (2009) 149, {\tt arXiv:0904.4601v2 [hep-th]}.

\bibitem{2c} D. Green, E. Silverstein, D. Starr, ``Attractor Explosions
and Catalyzed Vacuum Decay", Phys. Rev. D {\bf 74} (2006) 024004,
{\tt arXiv:hep-th/0605047v1}.

\bibitem{2d} A. Sen, ``Walls of Marginal Stability and Dyon Spectrum in N = 4
Supersymmetric String Theories", JHEP {\bf 0705} (2007) 039, {\tt
arXiv:hep-th/0702141v3}.

\bibitem{2e} S. Bellucci, S. Ferrara, R. Kallosh, A. Marrani, ``Extremal Black
Hole and Flux Vacua Attractors", Lect. Notes Phys. {\bf 755}
(2008) 115-191, {\tt arXiv:0711.4547v1 [hep-th]}.

\bibitem{2f} S. Ferrara, R. Kallosh, ``Supersymmetry and Attractors",
Phys. Rev. D {\bf 54} (1996) 1514-1524, {\tt
arXiv:hep-th/9602136v3}.

\bibitem{2g} S. Bellucci, S. Ferrara, M. Gunaydin, A. Marrani, ``SAM
Lectures on Extremal Black Holes in d = 4 Extended Supergravity",
Springer Proceedings in Phys. {\bf 134} (2010) 1-30, {\tt
arXiv:0905.3739 [hep-th]}.

\bibitem{2h} S. Bellucci, A. Marrani, E. Orazi, A. Shcherbakov, ``Attractors with Vanishing Central Charge",
Phys. Lett. B {\bf 655} (2007) 185-195, {\tt arXiv:0707.2730
[hep-th]}.

\bibitem{2i} S. Bellucci, S. Ferrara, A. Marrani, ``Supersymmetric mechanics.
Vol. 2: The attractor mechanism and space time singularities",
Lect. Notes Phys. {\bf 701} (2006) 1-225.

\bibitem{2l} S. Bellucci, S. Ferrara, A. Marrani, A. Shcherbakov,
``Splitting of Attractors in 1-modulus Quantum Corrected Special
Geometry", JHEP {\bf 0802} (2008) 088, {\tt arXiv:0710.3559
[hep-th]}.

\bibitem{2m} S. Bellucci, S. Ferrara, A. Shcherbakov, A. Yeranyan,
``Black hole entropy, flat directions and higher derivatives",
JHEP {\bf 0910} (2009) 024, {\tt arXiv:0906.4910 [hep-th]}.

\bibitem{3a} G. Villadoro, F. Zwirner, ``D terms from D-branes, gauge invariance
and moduli stabilization in flux compactifications", JHEP {\bf
0603} (2006) 087, {\tt arXiv:hep-th/0602120v2}.

\bibitem{3b} A. Achucarro, A. Celi, M. Esole, J. V. den Bergh,
A. V. Proeyen, ``D-term cosmic strings from N=2 Supergravity",
JHEP {\bf 0601} (2006) 102, {\tt arXiv:hep-th/0511001v2}.

\bibitem{3c} J. T. Liu, W. A. Sabra, ``Multi-centered black holes in gauged
D=5 supergravity", Phys. Lett. B {\bf 498} (2001) 123-130, {\tt
arXiv:hep-th/0010025v2}.

\bibitem{4a} S. Bellucci, S. Ferrara, A. Marrani, A. Yeranyan,
``d = 4 Black Hole Attractors in N = 2 Supergravity with
Fayet-Iliopoulos Terms", Phys. Rev. D {\bf 77} (2008) 085027, {\tt
arXiv:0802.0141v1 [hep-th]}.

\bibitem{4b} E. Dudas, Y. Mambrini, S. Pokorski, A. Romagnoni,
``Moduli stabilization with Fayet-Iliopoulos uplift", JHEP {\bf
0804} (2008) 015, {\tt arXiv:0711.4934v1 [hep-th]}.

\bibitem{5} S. Ferrara, R. Kallosh, A. Strominger, ``N=2 Extremal Black
Holes", Phys. Rev. D {\bf 52} (1995) 5412-5416, {\tt
arXiv:hep-th/9508072v3}.

\bibitem{6} A. Strominger,
``Macroscopic Entropy of N = 2 Extremal Black Holes",
Phys. Lett. B {\bf 383} (1996) 39-43, {\tt
arXiv:hep-th/9602111v3}.

\bibitem{7} S. Bellucci, S. Ferrara, A. Marrani, ``Attractor Horizon
Geometries of Extremal Black Holes", CERN-PH-TH/2007 025, {\tt
arXiv:hep-th/0702019v1}.

\bibitem{8} S. Bellucci, S. Ferrara, A. Marrani, ``Attractors in Black",
Fortsch. Phys. {\bf 56} (2008) 761-785, {\tt arXiv:0805.1310v1
[hep-th]}.

\bibitem{9} C. G. Callan, J. M. Maldacena, ``D-brane Approach to Black
Hole Quantum Mechanics", Nucl. Phys. B {\bf 472} (1996) 591-610,
{\tt arXiv:hep-th/9602043v2}.

\bibitem{10} G. L. Cardoso, B. de Wit, J. Kppeli, T. Mohaupt,
``Asymptotic degeneracy of dyonic N=4 string states and black hole
entropy", JHEP {\bf 0412} (2004) 075, {\tt
arXiv:hep-th/0412287v1}.

\bibitem{11a} Miranda C. N. Cheng, Erik P. Verlinde,
``Wall Crossing, Discrete Attractor Flow and Borcherds Algebra",
SIGMA {\bf 4} (2008) 068, {\tt arXiv:0806.2337v2 [hep-th]}.

\bibitem{11b} A. Misra, P. Shukla, ``Moduli Stabilization, Large-Volume dS
Minimum Without anti-D3-Branes, (Non-)Supersymmetric Black Hole
Attractors and Two-Parameter Swiss Cheese Calabi-Yau's", Nucl.
Phys. B {\bf 799} (2008) 165-198, {\tt arXiv:0707.0105v6
[hep-th]}.

\bibitem{11c} S. Nampuri, P. K. Tripathy, S. P. Trivedi, ``On The
Stability of Non-Supersymmetric Attractors in String Theory", JHEP
{\bf 0708} (2007) 054, {\tt arXiv:0705.4554v1 [hep-th]}.

\bibitem{11d} J. de Boer, M. C. N. Cheng, R. Dijkgraaf, J. Manschot,
E. Verlinde, ``A Farey Tail for Attractor Black Holes", JHEP {\bf
0611} (2006) 024, {\tt arXiv:hep-th/0608059v2}.

\bibitem{11e} B. Pioline, ``Lectures on on Black Holes, Topological Strings and
Quantum Attractors (2.0)", Class. Quant. Grav. {\bf 23} (2006)
S981, {\tt arXiv:hep-th/0607227v5}.

\bibitem{11f} S. Ferrara, R. Kallosh, ``On N=8 attractors", Phys. Rev. D {\bf
73} (2006) 125005, {\tt arXiv:hep-th/0603247v2}.

\bibitem{11g} M. Gunaydin, A. Neitzke, B. Pioline, A. Waldron, ``BPS
black holes, quantum attractor flows and automorphic forms", Phys.
Rev. D {\bf 73} (2006) 084019, {\tt arXiv:hep-th/0512296v2}.

\bibitem{11h} R. Kallosh, ``New Attractors", JHEP {\bf 0512} (2005)
022, {\tt arXiv:hep-th/0510024v3}.

\bibitem{11i} R. Kallosh, ``Flux vacua as supersymmetric attractors",
{\tt arXiv:hep-th/0509112v3}.

\bibitem{11j} H. Ooguri, A. Strominger, C. Vafa,
``Black Hole Attractors and the Topological String", Phys. Rev. D
{\bf 70} (2004) 106007, {\tt arXiv:hep-th/0405146v2}.

\bibitem{11k} G. Dvali, A. Vilenkin, ``Cosmic Attractors and Gauge
Hierarchy", Phys. Rev. D {\bf 70} (2004) 063501, {\tt
arXiv:hep-th/0304043v2}.

\bibitem{12a} F. Weinhold, ``Metric geometry of equilibrium thermodynamics", J. Chem.
Phys. {\bf 63} (1975) 2479, DOI:10.1063/1.431689.

\bibitem{12b} F. Weinhold, ``Metric geometry of equilibrium thermodynamics. II,
Scaling, homogeneity, and generalized GibbsDuhem relations", ibid
J. Chem. Phys {\bf 63} (1975) 2484.

\bibitem{12c} G. Ruppeiner, ``Riemannian geometry in thermodynamic fluctuation
theory", Rev. Mod. Phys {\bf 67} (1995) 605, [Erratum 68 (1996)
313].

\bibitem{12d} G. Ruppeiner, ``Thermodynamics: A Riemannian geometric model", Phys.
Rev. A {\bf 20} (1979) 1608.

\bibitem{12e} G. Ruppeiner, ``Thermodynamic Critical Fluctuation Theory?", Phys.
Rev. Lett. {\bf 50} (1983) 287.

\bibitem{12f} G. Ruppeiner, ``New thermodynamic fluctuation theory using path integrals",
Phys. Rev. A {\bf 27} (1983) 1116.

\bibitem{12g} G. Ruppeiner, C. Davis, ``Thermodynamic curvature of the
multi-component ideal gas", Phys. Rev. A {\bf 41} (1990) 2200.

\bibitem{12h} J. Shen, R. G. Cai, B. Wang, R. K. Su, ``Thermodynamic
Geometry and Critical Behavior of Black Holes", Int. J. Mod. Phys.
A {\bf 22} (2007) 11-27, {\tt arXiv:gr-qc/0512035v1}.

\bibitem{12i} J. E. Aman, I. Bengtsson, N. Pidokrajt, ``Flat Information
Geometries in Black Hole Thermodynamics", Gen. Rel. Grav. {\bf 38}
(2006) 1305-1315, {\tt arXiv:gr-qc/0601119v1}.

\bibitem{12j} J. E. Aman, I. Bengtsson, N. Pidokrajt, ``Geometry of black hole
thermodynamics", Gen. Rel. Grav. {\bf 35} (2003) 1733, {\tt
arXiv:gr-qc/0304015v1}.

\bibitem{12k} J. E. Aman, N. Pidokrajt, ``Geometry of Higher-Dimensional Black
Hole Thermodynamics", Phys. Rev. D {\bf 73} (2006) 024017, {\tt
arXiv:hep-th/0510139v3}.

\bibitem{12l} T. Sarkar, G. Sengupta, B. N. Tiwari, ``On the Thermodynamic
Geometry of BTZ Black Holes", JHEP {\bf 0611} (2006) 015, {\tt
arXiv:hep-th/0606084v1}.

\bibitem{12m} T. Sarkar, G. Sengupta, B. N. Tiwari,
``Thermodynamic Geometry and Extremal Black Holes in String
Theory", JHEP {\bf 0810} (2008) 076, {\tt arXiv:0806.3513v1
[hep-th]}.

\bibitem{12n} B. N. Tiwari, ``Sur les corrections de la g\'{e}eom\'{e}trie
thermodynamique des trous noirs", Quantum Gravity, Hoelback
(2008), arXiv:0801.4087v1 [hep-th].

\bibitem{12o} S. Bellucci, B. N. Tiwari, ``On the Microscopic Perspective of Black Branes Thermodynamic
Geometry", Entropy 2010, 12, 2097-2143, {\tt arXiv:0808.3921v1
[hep-th]}.

\bibitem{12oo} S. Bellucci, B. N. Tiwari, ``State-space Correlations and
Stabilities'', Phys. Rev. D {\bf 82} (2010) 084008, {\tt
arXiv:0910.5309v1 [hep-th]}.

\bibitem{12ooo} S. Bellucci, B. N. Tiwari,
``An Exact Fluctuating 1/2-BPS Configuration'', JHEP {\bf 1005}
(2010) 023, {\tt arXiv:0910.5314v2 [hep-th]}.

\bibitem{12o1} S. Bellucci, B. N. Tiwari,
``State-space Manifold and Rotating Black Holes'', {\tt
arXiv:1010.1427v1 [hep-th]}.

\bibitem{12o2} S. Bellucci, B. N. Tiwari,
``Black Strings, Black Rings and State-space Manifold'', {\tt
arXiv:1010.3832v1 [hep-th]}.

\bibitem{12o3} S. Bellucci, B. N. Tiwari,
``Thermodynamic Geometry and Hawking Radiation'', JHEP {\bf 1011}
(2010) 030, {\tt arXiv:1009.0633 [hep-th]}.

\bibitem{12p} S. Bellucci, V. Chandra, B. N. Tiwari, ``On the Thermodynamic
Geometry of Hot QCD", to appear in Int. J. Mod. Phys. A, {\tt
arXiv:0812.3792v1 [hep-th]}.

\bibitem{12p1} S. Bellucci, V. Chandra, B. N. Tiwari, ``Thermodynamic
Geometric Stability of Quarkonia states", {\tt arXiv:1010.4225v2
[hep-th]}.

\bibitem{12p2} S. Bellucci, V. Chandra, B. N. Tiwari, ``A geometric
approach to correlations and quark number susceptibilities", {\tt
arXiv:1010.4405v1 [hep-th]}.

\bibitem{12p3} S. Bellucci, B. N. Tiwari, ``Thermodynamic Geometry:
Evolution, Correlation and Phase Transition", {\tt
arXiv:1010.5148v2 [cond-mat.stat-mech]}.

\bibitem{12q} L. D. Landau, S. Lifshitz, ``Statistical Mechanics",
Pergaman Press (1980) Hungary.

\bibitem{12r} D. Tranah, P. T. Landsberg, ``Collective Phenomena",
{\bf 3} (1980) 81.

\bibitem{12s} G. Arcioni, E. Lozano-Tellechea, ``Stability and critical
phenomena of black holes and black rings", Phys. Rev. D {\bf 72}
(2205) 104021.

\bibitem{12t} J. E. Aman, N. Pidokrajt, ``Geometry of higher-dimensional black
hole thermodynamics", Phys. Rev. D {\bf 73} (2006) 024017.

\bibitem{12u}  G. Ruppeiner, ``Thermodynamic curvature and phase transitions in
Kerr-Newman black holes", Phy. Rev. D {\bf 78} (2008) 024016.

\bibitem{13a} E. Calabi, ``A construction of nonhomogeneous Einstein
metrics", Proc. of Symp. in Pure Mathematics, {\bf 27} (1975)
17-24, AMS, Providence.

\bibitem{13b} J. Li, S. T. Yau, ``Hermitian Yang-Mills connections on
non-K\"{a}hler manifolds", Mathematical aspects of string theory,
World Scientific (1987).

\bibitem{13c} A. Klemm, S. Theisen, ``Considerations of One-Modulus
Calabi-Yau Compactifications: Picard-Fuchs Equations, K\"{a}hler
Potentials and Mirror Maps", Nucl. Phys. B {\bf 389} (1993)
153-180, {\tt arXiv:hep-th/9205041v1}.

\bibitem{13d} P. S. Aspinwall, ``The Landau-Ginzburg to Calabi-Yau Dictionary
for D- Branes", J. Math. Phys. {\bf 48} (2007) 082304, {\tt
arXiv:hep-th/0610209v2}.

\bibitem{13e} A. Ricco, ``Brane superpotential and local Calabi-Yau
manifolds", Int. J. Mod. Phys. A {\bf 23} (2008) 2187-2189, {\tt
arXiv:0805.2738v1 [hep-th]}.

\bibitem{13f} A. Belhaj, ``On Black Objects in Type IIA Superstring Theory on
Calabi-Yau Manifolds", African Journal of Mathematical Physics,
{\bf 6} (2008) 49-54, {\tt arXiv:0809.1114v2 [hep-th]}.

\bibitem{14} A. Klemm, M. Kreuzer, E. Riegler, E. Scheidegger,
``Topological String Amplitudes, Complete Intersection Calabi-Yau
Spaces and Threshold Corrections", JHEP {\bf 0505} (2005) 023,
{\tt arXiv:hep-th/0410018v2}.

\bibitem{15} A. Klemm, M. Marino, ``Counting BPS states on the
Enriques Calabi-Yau", Commun. Math. Phys. {\bf 280} (2008) 27-76,
{\tt arXiv:hep-th/0512227v3}.

\bibitem{16} M. Aganagic, V. Bouchard, A. Klemm, ``Topological Strings
and (Almost) Modular Forms", Commun. Math. Phys. {\bf 277} (2008)
771-819, {\tt arXiv:hep-th/0607100v2}.

\bibitem{17a} K. Hotta, T. Kubota, ``Exact Solutions and the Attractor
Mechanism in Non-BPS Black Holes", Prog. Theor. Phys. {\bf 118}
(2007) 969-981, {\tt arXiv:0707.4554v3 [hep-th]}.

\bibitem{17b} M. P. G. del Moral, ``A New Mechanism of Kahler Moduli
Stabilization in Type IIB Theory", JHEP {\bf 0604} (2006) 022,
{\tt arXiv:hep-th/0506116v3}.

\bibitem{17c} G. Gibbons, R. Kallosh, B. Kol, ``Moduli, Scalar Charges,
and the First Law of Black Hole Thermodynamics", Phys. Rev. Lett.
{\bf 77} (1996) 4992-4995, {\tt arXiv:hep-th/9607108v2}.

\bibitem{18a} A. Sinha, N. V. Suryanarayana,
``Two-charge small black hole entropy:  String-loops and
multistrings", JHEP {\bf 0610} (2006) 034, {\tt
arXiv:hep-th/0606218v1}.

\bibitem{18b} A. Sinha, N. V. Suryanarayana, ``Extremal
single-charge small black holes: Entropy function analysis",
Class. Quant. Grav. {\bf 23} (2006) 3305, {\tt
arXiv:arXiv:hep-th/0601183v2}.

\bibitem{18c} A. Dabholkar, N. Iizuka, A. Iqubal, A. Sen, M. Shigemori,
``Spinning Strings as Small Black Rings", JHEP {\bf 0704} (2007)
017, {\tt arXiv:hep-th/0611166v2}.

\bibitem{18d} A. Dabholkar, A. Sen, S. Trivedi,
``Black Hole Microstates and Attractor Without Supersymmetry",
JHEP {\bf 0701} (2007) 09, {\tt arXiv:hep-th/0611143v2}.

\bibitem{18e} D. Astefanesei, K. Goldstein, R. P. Jena, A. Sen, S. P. Trivedi,
``Rotating Attractors", JHEP {\bf 0610} (2006) 058, {\tt
arXiv:hep-th/0606244v2}.

\bibitem{18f} D. Astefanesei, K. Goldstein, S. Mahapatra, ``Moduli and
(un)attractor black hole thermodynamics", {\tt
arXiv:hep-th/0611140v4}.

\bibitem{19a} A. Dabholkar, F. Denef, G. W. Moore, B. Pioline,
``Precision Counting of Small Black Holes", JHEP {\bf 0510} (2005)
096, {\tt arXiv:hep-th/0507014v1}.

\bibitem{19b} A. Sen, ``Stretching the Horizon of a Higher Dimensional
Small Black Hole", JHEP {\bf 0507} (2005) 073, {\tt
arXiv:hep-th/0505122v2}.

\bibitem{20a} A. D. Aleksandrov and V. A. Zalgaller,
``Intrinsic Geometry of Surfaces", {\bf 15E} (1967) AMS ISBN:
0-8218-3276-X.

\bibitem{20b} G. E. Bredon, ``Topology and Geometry", Graduate Texts in
Mathematics, {\bf 139} (1993) Springer.

\bibitem{20c} M. Atiyah, ``Complex Geometry and Analysis",
Lecture Notes in Mathematics, Springer Berlin/ Heidelberg, {\bf
1422} (1990) ISSN 0075-8434, Pages 1-13.

\bibitem{20d} J. Madore, S. Schraml, P. Schupp, J. Wess, ``External
Fields as Intrinsic Geometry", Eur. Phys. J. C {\bf 18} (2001)
785-794, {\tt arXiv:hep-th/0009230v1}.

\bibitem{21} C. Burrage, A. C. Davis, ``P-term potentials from 4-D
supergravity", JHEP {\bf 06} (2007) 086, {\tt arXiv:0705.1657v1
[hep-th]}.


\end{thebibliography}
\end{document}